\pdfoutput=1
\documentclass[cernpreprint, atlasdraft=false, texlive=2023, UKenglish, texmf, orcidlogo]{atlasdoc}
 
\usepackage[subfig=true]{atlaspackage}
\usepackage{xfrac}
\usepackage{atlasbiblatex}
\usepackage{tikz-feynman}
 
\usepackage[unit=false, hepparticle=true, hepprocess=true, jetetmiss=true, misc=true]{atlasphysics}
 
\graphicspath{{logos/}{figures/}}
 
\usepackage{ANA-EXOT-2020-07-PAPER-defs}

\AtlasTitle{Search for third-generation vector-like leptons
in $pp$ collisions at $\sqrt{s} = 13\,\text{TeV}$ with the ATLAS detector}
 
\AtlasAbstract{
A search for vector-like leptons
in multilepton (two, three, or four-or-more electrons plus muons) final states with zero or more
hadronic
$\tau$-lepton decays is presented. The search is performed using
a dataset corresponding to an integrated luminosity of 139~fb$^{-1}$
of proton--proton collisions at a centre-of-mass energy of
13~TeV recorded by the ATLAS detector at the LHC.
To maximize the separation of signal and background, a machine-learning classifier is used.
No excess of events is observed beyond the Standard Model expectation.
Using a doublet vector-like lepton model, vector-like leptons coupling
to third-generation Standard Model leptons are excluded
in the mass range from 130~GeV to 900~GeV
at the 95\% confidence level, while the highest excluded mass is expected
to be 970~GeV.
}
 
\author{The ATLAS Collaboration}
 
\AtlasJournalRef{JHEP 07 (2023) 118}
\AtlasDOI{10.1007/JHEP07(2023)118}
\PreprintIdNumber{CERN-EP-2022-236}

\hypersetup{pdftitle={ATLAS document},pdfauthor={The ATLAS Collaboration}}
 
\begin{document}
 
\maketitle
 
\tableofcontents
 
\section{Introduction}
\label{sec:intro}

The predictions of the Standard Model (SM) of particle physics are in excellent agreement with
measurements of \mbox{proton--proton} (\pp) collisions
at the Large Hadron Collider (LHC)~\cite{Evans:2008zzb}.
Nonetheless, some
aspects of our universe cannot be explained within the framework of the SM,
such as the excess of matter over antimatter, the origin of the neutrino masses,
the nature of dark matter and dark energy, and the hierarchy and fine-tuning
problems~\cite{Workman:2022ynf, Martin:1997ns}.
Many possible ways to find solutions have been proposed, including models based on supersymmetry,
which help to explain why the Higgs boson's mass is very far from the Planck scale.
However, the measured Higgs boson branching fractions are in good agreement
with the SM predictions, so it is not easy to accommodate new particles whose masses are
generated via the Higgs mechanism~\cite{Djouadi:2012ae, Eberhardt:2012gv, Lenz:2013iha}.
One class of particles that are motivated by a variety of phenomenological models
based on string theory or large extra dimensions are vector-like fermions that
transform as non-chiral representations of the unbroken SM
gauge group~\cite{Frampton:1999xi}. They therefore have Dirac masses and
decouple from the electroweak scale in the large-mass limit.
 
\begin{figure}[htbp]\centering
\subfloat[]{\label{fig:fyna}\centering
\includegraphics[width=0.32\textwidth]{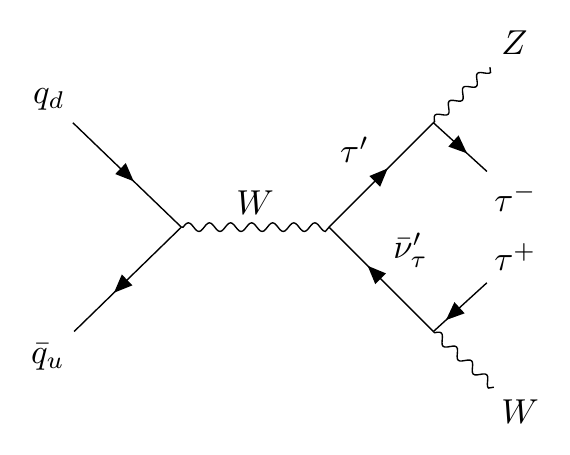}}\hfill
\subfloat[]{\label{fig:fynb}\centering
\includegraphics[width=0.32\textwidth]{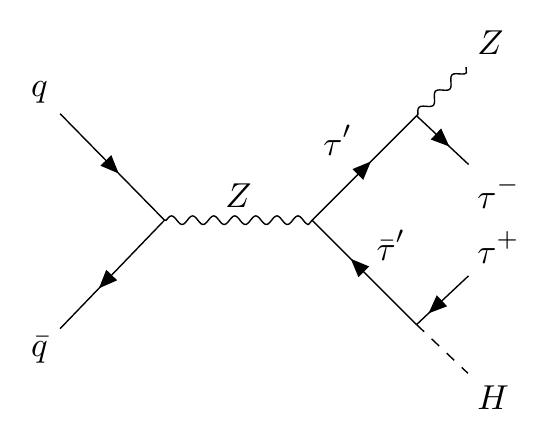}}\hfill
\subfloat[]{\label{fig:fync}\centering
\includegraphics[width=0.32\textwidth]{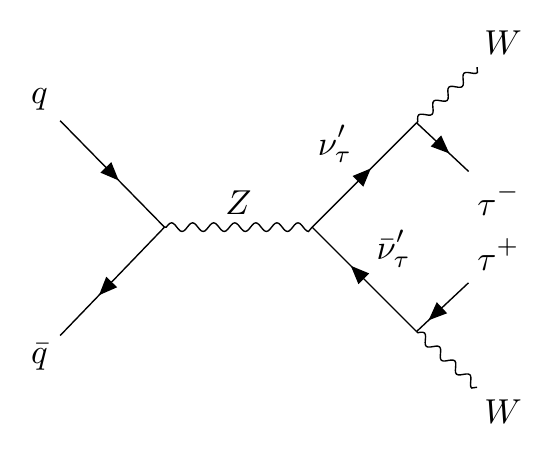}}
\caption{Examples of LO Feynman diagrams for VLL production and decay:
\protect\subref{fig:fyna} $\tau^{\prime}\bar{\nu}^{\prime}_{\tau}$ production followed by $\tau^{\prime}$ decay into $Z\tau$ and $\nu^{\prime}_{\tau}$ decay into $\tau^+W^-$, where $q_u$ ($q_d$) represents a weak isospin \sfrac{$+1$}{$2$} (\sfrac{$-1$}{$2$})
quark,
\protect\subref{fig:fynb} production of $\tau^{\prime}\bar{\tau}^\prime$ followed by the decay into $Z\tau^-$ and $\tau^+H$, and
\protect\subref{fig:fync} $\nu^{\prime}_{\tau}\bar{\nu}^{\prime}_{\tau}$ production followed by $\nu^{\prime}_{\tau}$ decay into $W\tau$.}
\label{fig:feynman_diagrams}
\end{figure}

A large number of searches for vector-like quarks have been performed at the
Tevatron~\cite{D0:2010ckq, CDF:2009gat} and the LHC~\cite{
CMS-EXO-11-005,
CMS-EXO-11-099,
EXOT-2011-22,
CMS-B2G-13-005,
CMS-B2G-13-006,
CMS-B2G-12-015,
CMS-B2G-14-002,
EXOT-2014-10,
EXOT-2013-18,
EXOT-2016-15,
CMS-B2G-16-024,
CMS-B2G-17-003,
CMS-B2G-17-011,
CMS-B2G-17-012,
EXOT-2016-14,
EXOT-2016-13,
EXOT-2017-34,
CMS-B2G-18-005}.
In addition, the CMS Collaboration has performed a search for vector-like
leptons (VLL) in the
$\Pgt$ plus three $b$-jet final state~\cite{CMS-B2G-21-004-a}.
The search was performed in the context
of the \enquote{4321~model}~\cite{4321_model,benchmark,nonuniversal4321}
over a mass range of \SIrange[range-units=single]{500}{1050}{\GeV}.
Following the
suggestion in Refs.~\cite{Kumar:2015tna,Bhattiprolu:2019vdu}, a search
for VLL in a doublet model has been performed by ATLAS and is presented in this
article. A similar search was performed by the CMS Collaboration
using \SI{138}{\ifb} of \pp collisions at $\sqrt{s}=\SI{13}{\TeV}$
and excludes vector-like \Pgt-lepton
masses below \SI{1045}{\GeV} at the 95\%
confidence level (CL)~\cite{CMS-EXO-21-002}.
The VLL doublet $L'=(\Pgngt',\Pgt')$ comprises two fermions of
approximately equal mass that couple only to the third-generation
leptons.
The VLL production cross section is dominated by $pp\to \Pgngt'\Pgt'$,
which is approximately 3.7 times greater than either
the $\pp\to\Pgt'^+\Pgt'^-$ or $pp\to\Pgngt'\Pagngt'$ modes,
which have approximately equal cross sections.
The $\Pgngt'$ decays exclusively into $\PWp\Pgtm$, while the
$\Pgt'$ decays are $\Pgt'^-\to \PZ\Pgt^-$ and $\Pgt'^-\to \PH\Pgt^-$, where
the branching fraction of the former is larger, but they asymptotically approach
each other with increasing $\tau'$ mass ($\Mtaup$) due to the Goldstone
equivalence principle~\cite{Cornwall:1974km}. Examples of
leading-order (LO) Feynman diagrams for VLL
production and decay are displayed in Figure~\ref{fig:feynman_diagrams}.
Given the possible
decays, the search is performed by selecting events containing at least
two charged light leptons,
$e^\pm$ or $\mu^\pm$, zero or more $\tau$-leptons decaying hadronically, and
a momentum imbalance transverse to the beam. To achieve
better background rejection than is possible with an event selection based
on kinematic and topological variables, a boosted decision tree (BDT)
algorithm is utilized as an event classifier~\cite{FREUND1997119, breiman2017classification}.
 
This article is organized as follows. A brief description of the
ATLAS detector is given in Section~\ref{sec:detector}.
Section~\ref{sec:datasim} presents the data and simulation samples used in
this search. The reconstruction of objects used
in the search for a VLL signal is delineated in Section~\ref{sec:object}.
Section~\ref{sec:analysis} describes techniques used to perform
the event selection, while Section~\ref{sec:bkg} outlines the method used
to estimate the backgrounds. A discussion of the systematic
uncertainties is given in Section~\ref{sec:syst}. The statistical method
used to arrive at the 95\% CL upper limit on the
VLL production cross section, and hence the mass exclusion region, is described
in Section~\ref{sec:results}. Finally, the analysis and results are
summarized in Section~\ref{sec:conclusion}.


\section{ATLAS detector}
\label{sec:detector}

\newcommand{\AtlasCoordFootnote}{
ATLAS uses a right-handed coordinate system with its origin at the nominal interaction point (IP)
in the centre of the detector and the \(z\)-axis along the beam pipe.
The \(x\)-axis points from the IP to the center of the LHC ring,
and the \(y\)-axis points upwards.
Cylindrical coordinates \((r,\phi)\) are used in the transverse plane,
\(\phi\) being the azimuthal angle around the \(z\)-axis.
The pseudorapidity is defined in terms of the polar angle \(\theta\) as \(\eta = -\ln \tan(\theta/2)\).
Angular distance is measured in units of \(\Delta R \equiv \sqrt{(\Delta\eta)^{2} + (\Delta\phi)^{2}}\).}

The ATLAS detector~\cite{PERF-2007-01} at the LHC covers nearly the entire solid angle around the collision point.\footnote{\AtlasCoordFootnote}
It consists of an inner tracking detector surrounded by a thin superconducting solenoid, electromagnetic and hadron calorimeters,
and a muon spectrometer incorporating three large superconducting air-core toroidal magnets.
 
The inner-detector system (ID) is immersed in a \SI{2}{\tesla} axial magnetic field
and provides charged-particle tracking in the range \(|\eta| < 2.5\).
The high-granularity silicon pixel detector covers the vertex region and
provides up to four measurements per track,
the first hit normally being in the insertable B-layer (IBL) installed before Run~2~\cite{ATLAS-TDR-19,PIX-2018-001}.
It is followed by the silicon microstrip tracker (SCT), which usually provides eight measurements per track.
These silicon detectors are complemented by the transition radiation tracker (TRT),
which enables radially extended track reconstruction up to \(|\eta| = 2.0\).
The TRT also provides electron identification information
based on the fraction of hits (typically 30 in total) above a higher energy-deposit threshold corresponding to transition radiation.
 
The calorimeter system covers the pseudorapidity range \(|\eta| < 4.9\).
Within the region \(|\eta|< 3.2\), electromagnetic calorimetry is provided by barrel and
endcap high-granularity lead/liquid-argon (LAr) calorimeters,
with an additional thin LAr presampler covering \(|\eta| < 1.8\)
to correct for energy loss in material upstream of the calorimeters.
Hadron calorimetry is provided by the steel/scintillator-tile calorimeter,
segmented into three barrel structures within \(|\eta| < 1.7\), and two copper/LAr hadron endcap calorimeters.
The solid angle coverage is completed with forward copper/LAr and tungsten/LAr calorimeter modules
optimized for electromagnetic and hadronic energy measurements, respectively.
 
The muon spectrometer (MS) comprises separate trigger and
high-precision tracking chambers measuring the deflection of muons in a magnetic field generated by the superconducting air-core toroidal magnets.
The field integral of the toroids ranges between \num{2.0} and \SI{6.0}{\tesla\metre}
across most of the detector.
A set of precision chambers covers the region \(|\eta| < 2.7\) with three layers of monitored drift tubes,
complemented by cathode-strip chambers in the forward region, where the
particle flux is highest.
The muon trigger system covers the range \(|\eta| < 2.4\) with resistive-plate chambers in the barrel, and thin-gap chambers in the endcap regions.
 
Interesting events are selected by the first-level trigger system implemented in custom hardware,
followed by selections made by algorithms implemented in software in the high-level trigger~\cite{TRIG-2016-01}.
The first-level trigger accepts events from the \SI{40}{\MHz} bunch crossings at a rate below \SI{100}{\kHz},
which the high-level trigger further reduces in order to record events to disk at about \SI{1}{\kHz}.
 
An extensive software suite~\cite{ATL-SOFT-PUB-2021-001} is used in data simulation, in the reconstruction
and analysis of real and simulated data, in detector operations, and in the trigger and data acquisition systems
of the experiment.


\section{Data and simulation samples}
\label{sec:datasim}

The data sample used in this article corresponds to an integrated luminosity of \lumi
of \pp collisions
at a centre-of-mass energy ($\sqrt{s}$) of $\SI{13}{\TeV}$ collected by the ATLAS detector during the
2015--2018 data-taking periods
after requiring stable LHC beams and that all detector subsystems were
operational~\cite{DAPR-2018-01}. The primary luminosity measurement was performed using the LUCID-2 detector~\cite{Avoni2018iuv}.
Candidate events are required to satisfy at least one of the dilepton triggers (\Pe{}\Pe, \Pgm{}\Pgm, \Pe{}\Pgm)~\cite{TRIG-2016-01,TRIG-2018-05,
TRIG-2018-01}.
These triggers have looser identification and isolation requirements than the single-lepton triggers but have comparable signal efficiency for events satisfying the analysis selection.
The lowest \pT threshold for the leading lepton range from
\SIrange[parse-numbers=false, range-units=single]{12}{22}{\GeV}
and for the sub-leading lepton it ranged from \SIrange[parse-numbers=false, range-units=single]{8}{17}{\GeV}.

To evaluate the effects of the detector resolution and acceptance on
the signal and background, and to estimate the SM backgrounds, simulated event samples were produced using
dedicated event generators. The detector response to the final-state particles was then modelled using a
$\GEANT$-based Monte Carlo (MC) detector simulation~\cite{Agostinelli:2002hh, SOFT-2010-01}.
The simulated data must account for the fact that significantly more than one inelastic \pp collision occurs per bunch
crossing, with
the average number ranging from \numrange{13}{38}
for the 2015--2018 data-taking periods,
respectively. Inelastic
collisions were simulated using \PYTHIA[8.186]~\cite{Sjostrand:2007gs} with the A3 set of tuned parameters~\cite{ATL-PHYS-PUB-2016-017} and the \NNPDF[2.3lo]~\cite{Ball:2012cx} set of parton distribution functions (PDFs), and overlaid on the
signal and background MC samples.
These simulated events were reweighted to match the conditions of the collision data, specifically the number of  additional \pp interactions (\pileup).
 
In order to optimize the signal selection and to estimate its acceptance and efficiency, a
simulated VLL sample was generated using the model described in
Refs.~\cite{Kumar:2015tna,Bhattiprolu:2019vdu}. This model was implemented at
LO, using \MGNLO[2.6.2]~\cite{Alwall:2014hca} with the \NNPDF[2.3lo] PDF to generate the
parton-level process, and selecting events with two, three, or four-or-more light leptons ($\ell$) and
$\ge 0$ hadronically decaying $\tau$-leptons, with all leptons having
$\pT>\SI{18}{\GeV}$ and $\abseta<2.8$. The generated \tauvll mass points
are \SI{130}{\GeV} and \SIrange[range-units=single]{200}{1300}{\GeV} in steps of \SI{100}{\GeV}.
These samples were then reweighted to next-to-leading order (NLO) in QCD.
To perform the fragmentation and hadronization, these events were
processed using \Pythia[8.212] with the A14 set of tuned parameters~\cite{ATL-PHYS-PUB-2016-017}
and the \NNPDF[2.3lo] PDFs.
 
Since each VLL decays into a $\tau$-lepton and either a vector boson or a Higgs boson, the selection
requirements are $\ge 2$ light leptons and $\ge 0$ $\tau$-leptons decaying to hadrons
(\tauhad). The leptonic decays of  $\tau$-leptons are not explicitly considered in the analysis. 
Top quark and multi-vector-boson events can contribute to the background since they have similar final states.
Therefore, to estimate the SM backgrounds, samples of simulated events were generated, containing multiple
final-state leptons that result from the decays of either directly produced vector bosons or vector bosons from top quark
or \PH decays. Backgrounds from misidentified leptons are estimated with
data-driven methods, as discussed in Section~\ref{sec:bkg}.
 
The vector-boson samples were produced using
either \SHERPA[2.2.1] or \SHERPA[2.2.2]~\cite{Bothmann:2019yzt} with the \NNPDF[3.0nnlo]~\cite{Ball:2014uwa}
set of PDFs, and the set of tuned parton-shower parameters developed
by the authors of \SHERPA~\cite{Schumann:2007mg}.
The \Vjets samples, where $V$ is either a \Wboson or \Zboson boson, and the
semileptonic diboson samples were produced using \SHERPA[2.2.1], while the
fully leptonic diboson samples and the triboson samples were produced using
\SHERPA[2.2.2]. The matrix element (ME) calculations were matched and merged
with the \SHERPA parton shower using the \MEPSatNLO
prescription~\cite{Hoeche:2011fd,Hoeche:2012yf,Catani:2001cc,Hoeche:2009rj} based on the
Catani--Seymour dipole factorization procedure~\cite{Winter:2003tt}.
For the \Vjets samples, the calculation was performed at
NLO in QCD for up to two partons and at LO for up to four partons using the
\COMIX~\cite{Gleisberg:2008fv} and \OPENLOOPS~\cite{Buccioni:2019sur,Cascioli:2011va,Denner:2016kdg} libraries.
For the diboson sample, this was performed for one parton at NLO and up to three partons
at LO, while, for the triboson sample, the calculation was performed at
NLO for the inclusive sample and at LO for up to two partons.

The electroweak production of single-top events, and also $\ttbar$ and \ttH  production, was simulated by
\POWHEGBOX[v2]~\cite{Alioli:2010xd,Nason:2004rx,Frixione:2007vw} at NLO
with the \hdamp parameter\footnote{The
\hdamp parameter is a resummation damping factor and one of the
parameters that controls the matching of \POWHEG matrix elements to
the parton shower. It effectively regulates the
high-\pt radiation against which the \ttbar system recoils.} set to
\SI{1.5}{$m_{\text{top}}$}~\cite{ATL-PHYS-PUB-2016-020},
the \NNPDF[3.0nlo] ~\cite{Ball:2014uwa} PDF set, and the top-quark mass set to \SI{172.5}{\GeV}. The diagram removal scheme~\cite{Frixione:2008yi} was employed to handle interference between single-top $t+W$ associated production and \ttbar production~\cite{ATL-PHYS-PUB-2016-020}.
Partons were hadronized and showered by
\PYTHIA[8.230]~\cite{Sjostrand:2014zea}, using the A14 tune and the
\NNPDF[2.3lo] PDF set.

The \ttV, \tZ, \tWZ, $\Pqt{}\Pqt{}\Pqt$, \tttt and \ttWW events were generated by \MGNLO[2.3.3]~\cite{Alwall:2014hca} at NLO,
using the \NNPDF[3.0nlo] PDF set for \ttV, \tZ, \tWZ, $\Pqt{}\Pqt{}\Pqt$ and \ttWW, and the \NNPDF[3.1nlo] PDF set for \tttt.
Parton hadronization
and showering was performed by
\PYTHIA[8.210] using the A14 tune and the \NNPDF[2.3lo] PDF set.
A summary of simulated signal and background samples is provided in Table~\ref{tab:samples}.
 
\begin{table}[htbp]
\caption{Simulated signal and background samples used in this analysis. The physics process is listed, along with the ME generator and order of the ME calculation, the parton hadronization and showering model, and the PDF used to simulate the process.}
\centering
\begin{tabular}{l|llll}
\toprule
Process&ME Generator & ME Order &Showering Model&PDF \\
\toprule
Signal&\MGNLO&LO&\PYTHIA[8] & \NNPDF[2.3lo]\\
\midrule
$Z\rightarrow \ell^+\ell^-$&\SHERPA[2.2.1]&NNLO&\SHERPA&\NNPDF[3.0nnlo] \\
$W\rightarrow \ell\nu_\ell$&\SHERPA[2.2.1]&NNLO&\SHERPA&\NNPDF[3.0nnlo] \\
\midrule
$VV$ (lep)&\SHERPA[2.2.2]&NLO&\SHERPA& \NNPDF[3.0nnlo] \\
$VV$ (semi-lep)&\SHERPA[2.2.1]&NLO&\SHERPA&\NNPDF[3.0nnlo] \\
$VVV$&\SHERPA[2.2.2]&NLO&\SHERPA& \NNPDF[3.0nnlo]\\
\midrule
$t$&\POWHEGBOX[v2]&NLO&\PYTHIA[8]& \NNPDF[2.3lo] \\
$t\bar{t}$&\POWHEGBOX[v2]&NLO&\PYTHIA[8]& \NNPDF[2.3lo] \\
$t\bar{t}+WW$&\MGNLO&NLO&\PYTHIA[8]& \NNPDF[2.3lo] \\
$t\bar{t}+H$&\POWHEGBOX[v2]&NLO&\PYTHIA[8]&  \NNPDF[2.3lo] \\
$t\bar{t}+V$&\MGNLO&NLO&\PYTHIA[8]&  \NNPDF[2.3lo] \\
$ttt$&\MGNLO&NLO&\PYTHIA[8]& \NNPDF[2.3lo] \\
$t\bar{t}t\bar{t}$&\MGNLO&NLO&\PYTHIA[8]& \NNPDF[2.3lo] \\
$t+Z$&\MGNLO&NLO&\PYTHIA[8]& \NNPDF[2.3lo] \\
$t+WZ$&\MGNLO&NLO&\PYTHIA[8]& \NNPDF[2.3lo] \\
\bottomrule
\end{tabular}
\label{tab:samples}
\end{table}


\section{Object reconstruction}
\label{sec:object}

All events used in this analysis are required to contain a primary
vertex~\cite{ATL-PHYS-PUB-2015-026}. It is selected as the \pp collision vertex
candidate with the highest sum of the squared transverse momenta of all associated
tracks with $\pT > \SI{500}{\MeV}$. In addition, there must be at least
two tracks associated with that vertex.

Electron candidates are reconstructed from energy clusters in the
EM calorimeter that match a reconstructed track~\cite{EGAM-2018-01}.
They are required to have $\pT>\SI{30}{\GeV}$
in order to suppress the number of fake electrons.
In addition, they are required to be within $\abseta<2.47$ with
the region $1.37<\abseta<1.52$ being excluded since it
contains a significant amount of non-sensitive material in front of the calorimeter.
To match these objects to the primary vertex, the track's transverse and longitudinal
impact parameters ($d_0$ and $z_0$, respectively) are required to satisfy
$|d_0|/\sigma(d_0)<5$ and $|z_0\sin\theta|<\SI{0.5}{\milli \meter}$.
Furthermore, through the application of
one of several working points,
these candidates must satisfy object identification criteria~\cite{EGAM-2018-01}.
Each working point offers a trade-off between
identification efficiency and misidentification rate.
A likelihood-based discriminant is constructed from a set of
variables that enhance electron selection, while
rejecting photon conversions and hadrons misidentified as
electrons~\cite{EGAM-2018-01}.
An $\eta$- and \ET-dependent
selection is applied to the likelihood discriminant in order to define specific working points.
For this search, the \textit{Tight} likelihood working point is used, which
has a 75\% efficiency at $\ET=\SI{30}{\GeV}$, increasing to 88\% at $\ET=\SI{100}{\GeV}$~\cite{EGAM-2018-01},
when used to identify electrons from \PZ{}-boson decays.
Electrons are also required to be isolated using criteria based on ID tracks and topological clusters in the calorimeter; the \textit{Loose} isolation working point is applied and has an efficiency of approximately 99\%~\cite{EGAM-2018-01}.
Correction factors are applied to simulated electrons to take into account the small differences in
reconstruction, identification and isolation efficiencies between data and MC simulation.
 
Muon candidates are reconstructed
by combining a reconstructed track from the ID with one
from the muon spectrometer~\cite{MUON-2018-03}, with the requirement that $\pt>\SI{20}{\GeV}$
and $\abseta<2.5$.
In addition,
the transverse and longitudinal impact parameters of the track are required to satisfy
$|d_0|/\sigma(d_0)<3$ and $|z_0\sin\theta|<\SI{0.5}{\milli \meter}$.
To reject misidentified
muon candidates, primarily originating from pion and kaon decays, several
quality requirements are imposed on the muon candidate. The \textit{Medium}
working point is used to select muons with $\pT<\SI{300}{\GeV}$ and the \textit{HighPt}
working point is used for those with $\pT>\SI{300}{\GeV}$~\cite{MUON-2018-03}.
This choice ensures a 97\% efficiency at $\pT=\SI{30}{\GeV}$ for the \textit{Medium}
working point, and 76\% at $\pT=\SI{500}{\GeV}$ for the \textit{HighPt} working point
~\cite{MUON-2018-03}.
An isolation requirement based on ID tracks and topological clusters in the calorimeter
is imposed. The \textit{TightTrackOnly} isolation working point is used,
resulting in an efficiency between 94\% and 99\% for muons
from \PW-boson decays in simulated \ttbar events~\cite{MUON-2018-03}.
Similarly to electrons, correction factors are applied to simulated muons to account for the small differences between data and simulation.
 
Particle-flow (PFlow) jets within $\abseta<4.5$ are reconstructed using the \antikt algorithm~\cite{Cacciari:2008gp, Cacciari:2005hq}
with a radius parameter $R=0.4$~\cite{Fastjet},
using neutral PFlow constituents and charged constituents associated
with the primary vertex as input~\cite{PERF-2015-09}.
These jets are then calibrated to the particle level by applying a jet energy
scale derived from simulation~\cite{JETM-2018-05}. Furthermore, in situ corrections based on the collected
data are applied~\cite{JETM-2018-05}. A cleaning procedure is used to identify and remove
jets arising from calorimeter noise or non-collision backgrounds. To suppress jets arising from \pileup, a discriminant
called the \enquote{jet vertex tagger} (JVT) is constructed using a two-dimensional
likelihood method~\cite{PERF-2014-03}. The \textit{Medium} JVT working point is used, which
has an average efficiency of 95\%. Jets used in this analysis are required to have $\pt>\SI{20}{\GeV}$
and $\abseta<2.5$. Jets originating from \Pqb-quarks are identified using the DL1r \Pqb-tagging algorithm~\cite{FTAG-2018-01,FTAG-2019-07}.
The working point used corresponds to a \Pqb-tagging efficiency of 77\%~\cite{FTAG-2018-01,FTAG-2019-07},
measured in a sample of simulated $t\bar{t}$ events. The corresponding rejection factors are
approximately 130, 5, and 14 for light-quark and gluon jets, \Pqc-jets, and hadronically decaying
\Pgt-leptons, respectively. Correction factors are applied to the simulated jets to take account of the small differences in
reconstruction and identification efficiencies, and the energy scale and resolution differences, between data and MC simulation.

The event selection for this analysis considers only those
\Pgt-leptons that decay into final states containing
a $\nu_\tau$-neutrino and one-or-more hadrons, denoted by \tauhad. Since the $\nu_\tau$ escapes the detector volume
undetected, only the hadronic decay products consisting of one or three
charged hadrons and up to two neutral pions are visible and they are denoted by \tauhadvis.
Reconstruction of \tauhadvis~\cite{ATLAS-CONF-2017-029,ATL-PHYS-PUB-2015-045}
is seeded by jets reconstructed from topological clusters by the \antikt algorithm
with the radius parameter $R$ set to $0.4$.
The tracks associated with the jet are required to originate from the primary vertex by satisfying the impact parameter
requirements $|d_0|<\SI{1}{\milli \meter}$ and $|z_0\sin\theta|<\SI{1.5}{\milli \meter}$. If these requirements are satisfied,
the tracks are then required to be within a cone of
size $\DeltaR=0.2$ around the jet axis, surrounded by a conical isolation region covering $0.2<\DeltaR<0.4$, in order to be considered a \tauhadvis candidate.
The direction of the \tauhadvis candidate in ($\eta,\, \phi$) is calculated as the vector
sum of the topological clusters within $\DeltaR=0.2$ of the jet axis, using
the \tauhadvis vertex as the origin. A multivariate discriminant is used to
select tracks that were produced by charged \tauhad decay
products~\cite{ATLAS-CONF-2017-029,ATL-PHYS-PUB-2015-045}.
Reconstructed \tauhadvis objects are selected for the analysis if they
have exactly one or three associated tracks (1- or 3-prong) with a total charge equal to $\pm1$.
The \tauhadvis objects must also satisfy the requirements $\pT>\SI{20}{\GeV}$ and $\abseta<2.47$, excluding the region $1.37 < \abseta < 1.52$.
These requirements have an efficiency of about 85\% for 1-prong and 70\%
for 3-prong \tauhadvis objects~\cite{ATLAS-CONF-2017-029,ATL-PHYS-PUB-2015-045}
as estimated from simulated $\PZ/\Pggx\to\tautau$ events.
A multivariate regression technique trained on MC samples is used to determine
the \tauhadvis energy scale using information from associated tracks, calorimeter energy
clusters, and reconstructed neutral pions~\cite{ATLAS-CONF-2017-029,ATL-PHYS-PUB-2015-045} .
 
A recurrent neural network (RNN) classifier~\cite{ATL-PHYS-PUB-2019-033}
is employed to select \tauhad-initiated jets and reject those initiated by quarks
or gluons. The RNN is trained on simulated $\PZ/\Pggx\to\tautau$ (for signal) and
simulated dijet events (for background). The training variables are single-track variables,
and reconstructed kinematic and topological variables. This analysis uses
the \textit{Medium} working point with an efficiency of $75\%$ ($60\%$) for 1-prong (3-prong)
candidates and a background rejection factor of 35 (240).
A boosted decision tree~\cite{ATLAS-CONF-2017-029,ATL-PHYS-PUB-2015-045}
is used to reject electron backgrounds misidentified as 1-prong \tauhadvis objects.
Variables used for its training include information from the calorimeter, the tracking detector,
and the visible momentum measured from the reconstructed tracks.
The \textit{Tight} working point with an efficiency of $75\%$ is used. To assess the contribution from
misidentified leptons (Section~\ref{sec:bkg}), less stringent (Loose) object identification
requirements are applied. When selecting Loose \tauhadvis there is an additional
requirement for the RNN score to be at least 0.01.
In addition, a dedicated muon-veto criterion
is used to reject muons reconstructed as \tauhadvis. Correction factors are applied to
simulated \tauhad objects to take into account the small differences in
reconstruction and identification efficiencies between data and MC simulation.
The energy scale and resolution differences between data and MC simulation are also
accounted for by applying scale factors. A summary of the lepton object definitions is presented in Table~\ref{tab:object}.

\begin{table}[htbp]
\caption{Object definitions used in the analysis for leptons. Tight objects are those selected for the nominal analysis and pass the prescribed selection requirements. Loose objects are those used to assess the contribution from the fake-lepton background and fail one or more selection requirements.}
\begin{center}
\small
\begin{tabular}{l|l|l|l|l|l}
\toprule
&\multicolumn{2}{l|}{Electrons}&\multicolumn{2}{l|}{Muons}&\tauhadvis\\
\midrule
& \textit{Tight} likelihood &\textit{Loose} isolation &\textit{Medium} (\textit{HighPt})&\textit{TightTrackOnly} isolation&\textit{Medium} RNN\\
\midrule
Tight &Pass&Pass&Pass&Pass&Pass\\
Loose&Fail*&Fail*&Pass&Fail&Fail\\
\bottomrule
\multicolumn{6}{l}{*Loose electrons are defined as failing either the \textit{Tight} likelihood or the \textit{Loose} isolation requirements, but not both.}
\end{tabular}
\label{tab:object}
\end{center}
\end{table}
 
The missing transverse momentum (with magnitude \MET) \cite{ATLAS-CONF-2018-023} is
reconstructed as the negative vector sum of the $\pT$ of all the selected electrons, muons,
jets, and \tauhadvis. An extra track-based \enquote{soft term} is built using additional tracks
associated with the primary vertex, but not with any reconstructed object. The use
of this track-based soft term is motivated by improved performance in \MET in a high \pileup environment.
 
To avoid cases where the detector response to a single physical object is reconstructed as
two separate final-state objects, an overlap removal procedure is used.
If electron and muon candidates share a track, the electron candidate is removed.
After that, if the $\DeltaR_{y,\phi}$ distance\footnote{$\DeltaR_{y,\phi}$ is the Lorentz-invariant distance in the rapidity--azimuthal-angle plane, defined as $\DeltaR_{y,\phi} =\sqrt{(\Delta y)^2+(\Delta\phi)^2}$ where the rapidity is $y=(1/2)[(E+p_z)/(E-p_z)]$.}
between a jet and an electron candidate is less than 0.2, the jet is discarded.
If multiple jets satisfy this requirement, only the closest jet is removed.
For jet--electron distances between 0.2 and 0.4, the electron candidate is removed.
If the distance
between a jet and a muon candidate is less than 0.4, the muon candidate is removed if the jet has more
than two associated tracks; otherwise the jet is removed.
The \tauhad candidates are seeded from jets, so this procedure removes any ambiguity in their selection.


\section{Analysis strategy}
\label{sec:analysis}

Based on the production and decay modes of the \tauvll and \nuvll
given in Section~\ref{sec:intro}, the multilepton final states are expected to maximize
the signal sensitivity, and hence the multilepton final states
are used to search for the VLL signal.
To further optimize the signal sensitivity, the different decay modes are targeted
by splitting the data into seven training regions
(given in Table~\ref{tab:trainingBDT}) based on the numbers of light leptons
and \tauhad, and a \MET requirement. In all regions at least one
jet is required. Additional requirements are derived by
maximizing the signal significance through the
application of a BDT~\cite{FREUND1997119, breiman2017classification}.
This leads to the seven signal regions (SRs) defined in Table~\ref{tab:SR}.
\begin{table}[h!]
\caption{The regions used in the training and optimization of the BDT.
The training regions are split according to the number of light leptons ($N_{\ell}$),
and their charge (SS same charge, OS opposite charge) and flavour (SF same flavour,
OF opposite flavour), the number of \tauhad ($N_{\Pgt}$), and the missing transverse momentum (\MET).
Every region has a requirement that it contain at least one jet ($\Njet>0$).}
\label{tab:trainingBDT}
\centering
\resizebox{\textwidth}{!}{
\begin{tabular}{l|ccccccc}
\toprule
Variables&\multicolumn{7}{c}{BDT Training Regions}\\
\midrule
BDT &2$\ell$ SSSF, 1$\tau$&2$\ell$ SSOF, 1$\tau$&2$\ell$ OSSF, 1$\tau$&2$\ell$ OSOF, 1$\tau$&2$\ell$, $\geq$2$\tau$&3$\ell$, $\geq$1$\tau$&4$\ell$, $\geq$0$\tau$\\
\midrule
$N_{\ell}$&$2$&$2$&$2$&$2$&~~~\,$2$&~~~\,$3$&$\geq4$\\
 
Charge/flavour&SSSF&SSOF&OSSF&OSOF&-&-&-\\
 
$N_{\Pgt}$&$1$&$1$&$1$&$1$&$\geq2$&$\geq1$&$\geq0$\\
 
$\MET$ [GeV]&$\geq120$&$\geq90$&$\geq60$&$\geq100$&$\geq60$&$\geq90$&$\geq60$\\
\bottomrule
\end{tabular}
}
\end{table}

\begin{table}[h!]
\caption{The BDT score requirement imposed on the training regions that are used to define the signal regions. No other requirement differs between the two regions.}
\label{tab:SR}
\centering
\resizebox{\textwidth}{!}{
\begin{tabular}{l|ccccccc}
\toprule
Variables&\multicolumn{7}{c}{Signal Regions}\\
\midrule
BDT &2$\ell$ SSSF, 1$\tau$&2$\ell$ SSOF, 1$\tau$&2$\ell$ OSSF, 1$\tau$&2$\ell$ OSOF, 1$\tau$&2$\ell$, $\geq$2$\tau$&3$\ell$, $\geq$1$\tau$&4$\ell$, $\geq$0$\tau$\\
BDT Score&$\geq0.15$&$\geq0.1$&$\geq0.1$&$\geq0.1$&$\geq-0.11$&$\geq0.08$&$\geq0.08$\\
\bottomrule
\end{tabular}
}
\end{table}

\begin{figure}[htbp]
\centering
\subfloat[]{\label{fig:SubCRb}
\includegraphics[scale = 0.26]{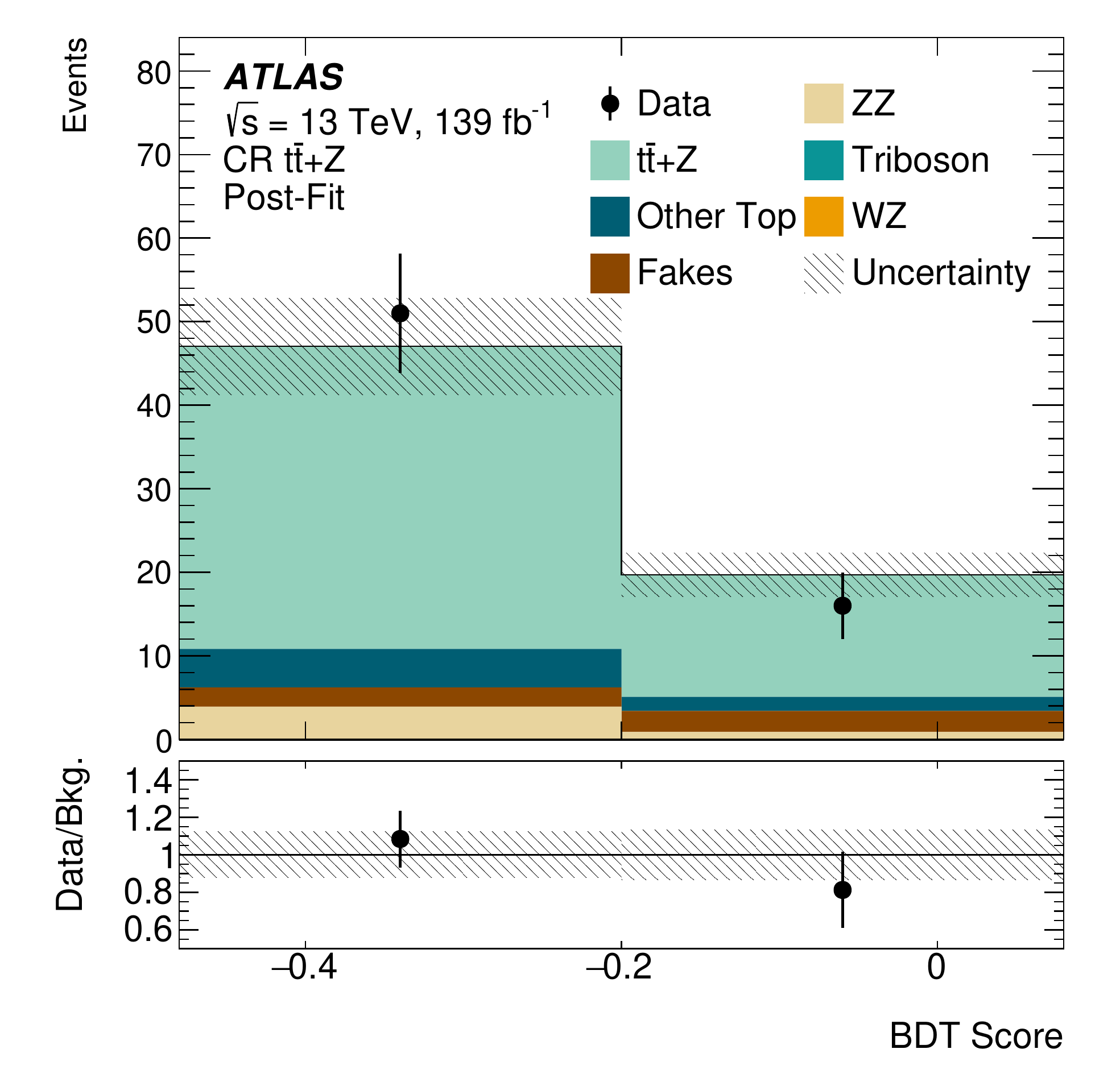}
}
\subfloat[]{\label{fig:SubCRa}
\includegraphics[scale = 0.26]{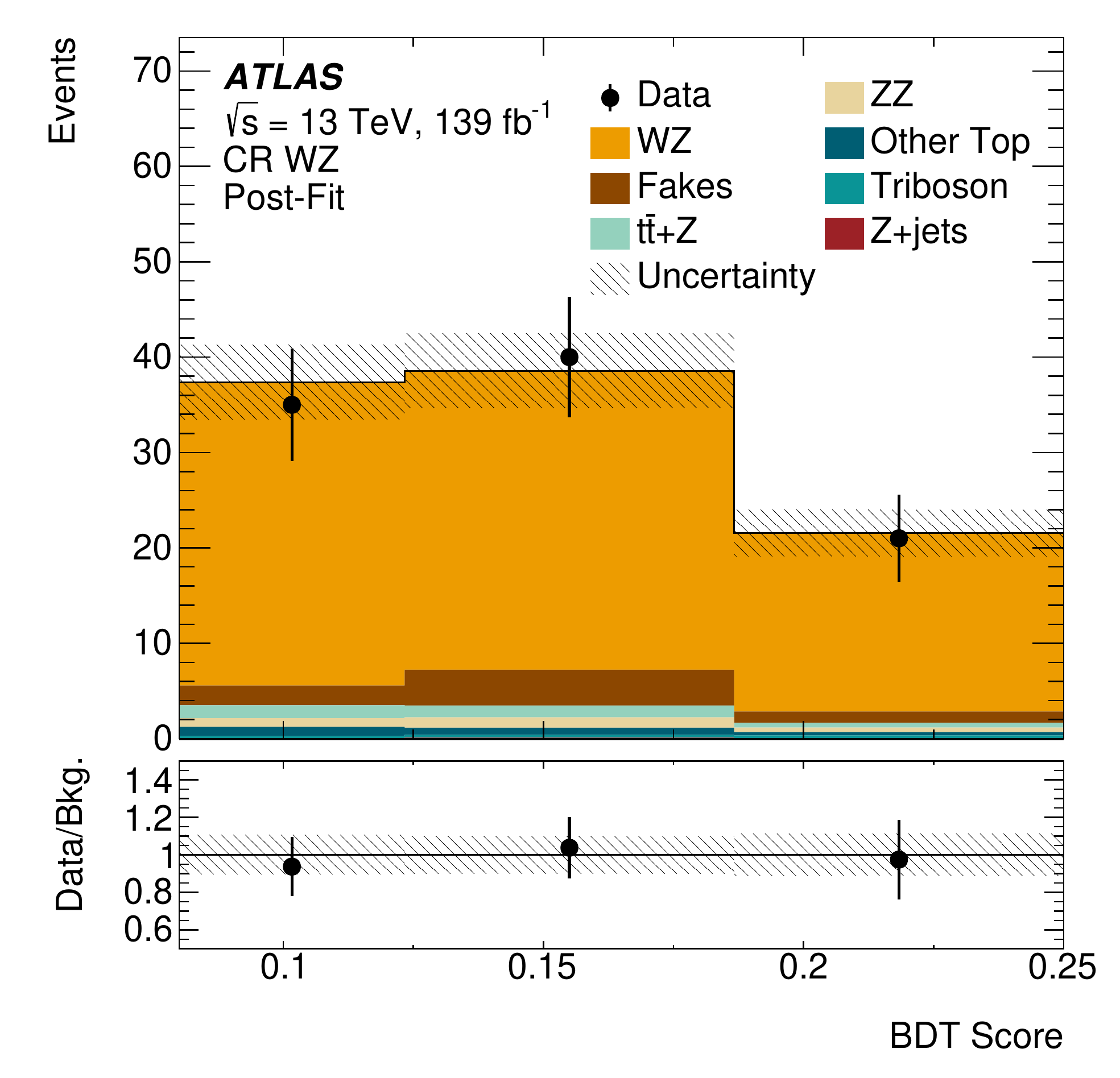}
}\subfloat[]{\label{fig:SubCRc}
\includegraphics[scale = 0.26]{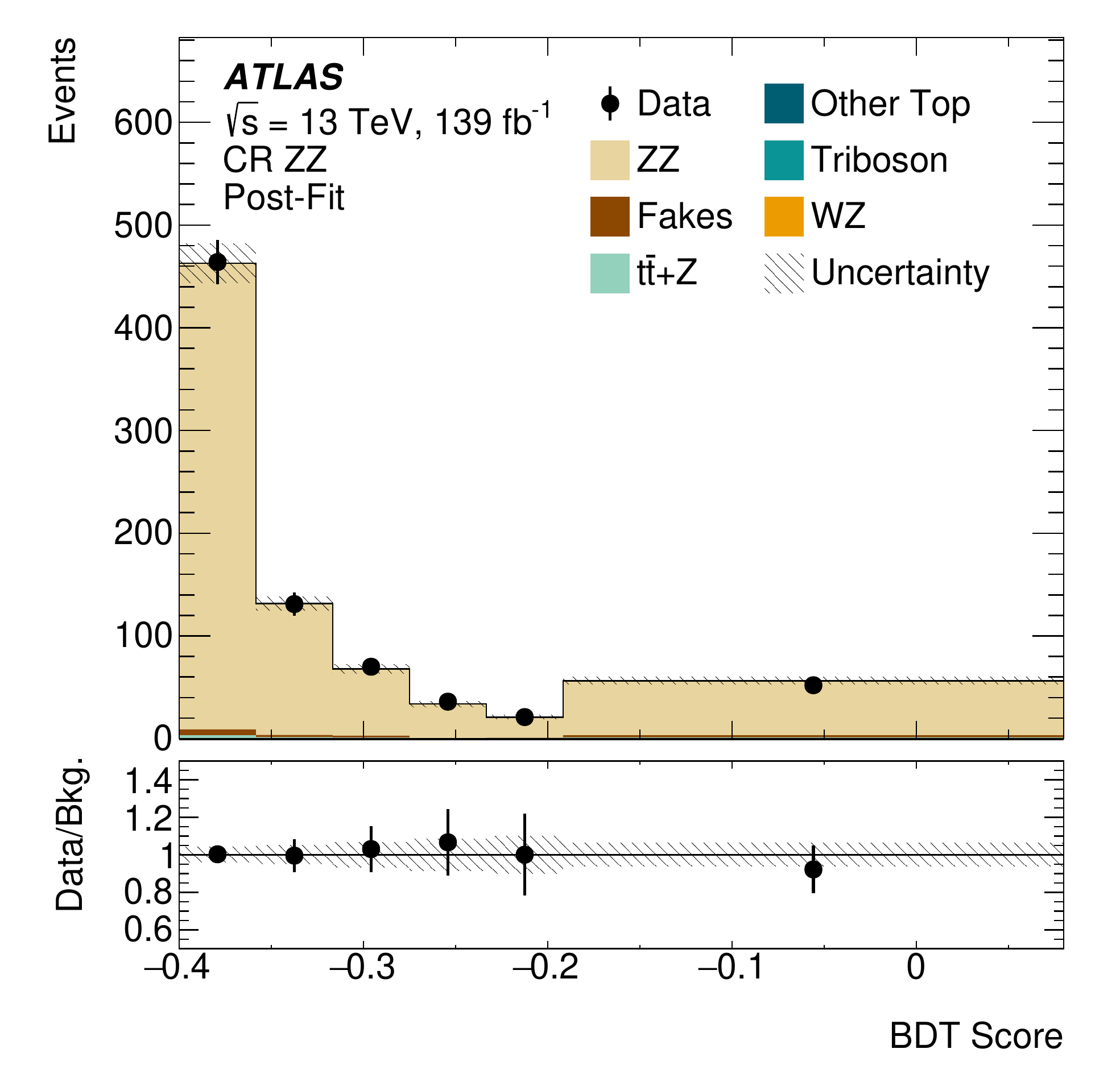}
}
\caption{A comparison of the BDT scores for data and the predicted background after performing the fit described in Section~\ref{sec:results}
in the CRs used to estimate the three dominant backgrounds:
\protect\subref{fig:SubCRb} \ttZ,
\protect\subref{fig:SubCRa} \PW{}\PZ, and
\protect\subref{fig:SubCRc} \PZ{}\PZ.
The first bin of the \ttZ and \PZ{}\PZ distributions
contains underflow events, while the last bin of the \PW{}\PZ
distribution contains overflow events.
The hatched band represents the combined statistical and systematic uncertainties.
\label{fig:CR}
}
\end{figure}

In addition to the seven SRs, three control regions (CRs) are used
in order to normalize the dominant physics background ($\ttbar+\PZ{}$, $\PW{}\PZ$, and $\PZ{}\PZ$)
estimates to data, and a fourth CR
is used to assess fake \tauhad objects originating from gluon-initiated jets and \pileup.
These CRs are defined in Table~\ref{tab:CR}.
Since events in the $WZ$ CR do not have a $\tau$-lepton,
its kinematic variables are set to zero when calculating the BDT score.
To confirm that the CRs are modelled correctly and that the obtained background normalization
factors are also valid in the regions with different numbers of light leptons and $\tau$-leptons, three validation
regions (VRs) are defined, as also shown in Table~\ref{tab:CR}.
An additional seven VRs are used
to confirm that the BDT models the data correctly. These VRs
are selected where the BDT distribution is expected to primarily contain background
events, as shown in Figure~\ref{fig:BDTdistros}.
The SRs, CRs, and VRs
are selected such that there is no overlap of events between the regions.
The BDT distribution shape in the CRs is shown in Figure~\ref{fig:CR},
which demonstrates good agreement between data and the background simulation.

\begin{table}[htbp]
\caption{The definition of the CRs used to determine the normalization
of the largest backgrounds, as well as the CR used to assess fake $\tau$-leptons originating from gluon-initiated jets and \pileup. In addition, the VRs used to validate
the CRs are also defined. Both the CRs and VRs
are selected so that they do not overlap with the SRs or with each other,
but are similar
enough to avoid problems when extrapolating between the regions.
Like the SRs, the CRs and VRs have the requirement that $\Njet>0$.
A dash indicates that the variable is not used for the
corresponding CR or VR.
}
\centering
\begin{tabular}{l|cccc|ccc}
\toprule
&\multicolumn{4}{c|}{Control Regions}&\multicolumn{3}{c}{Validation Regions}\\
&$t\bar{t}$+\PZ{}&\PW{}\PZ&\PZ{}\PZ&Fake \tauhad
&$t\bar{t}$+\PZ{}&\PW{}\PZ&\PZ{}\PZ\\
\midrule
BDT &4$\ell$, $\geq$0$\tau$ & 3$\ell$, $\geq$1$\tau$ & 4$\ell$, $\geq$0$\tau$ & 2$\ell$ OSSF, 1$\tau$ & 3$\ell$, $\geq$1$\tau$ & 2$\ell$ SSOF, 1$\tau$&3$\ell$, $\geq$1$\tau$\\
$N_{\ell}$&$\geq4$&$3$&$\geq4$&2&~~~\,3&2&~~~\,3\\
 
$N_{\tau}$&$\geq0$&$0$&$\geq0$&$1$ &$\geq 1$& 1&$\geq 1$\\
 
$N_{b}$&$>0$&$0$&~~~\,$0$&- &$>0$&0&~~~\,0\\
 
$\MET$ [GeV]&$\geq60$&$\geq90$&$<60$&$\geq60$&-&-&-\\
 
Charge/flavour &-&-&- &OSSF&- &SSOF&-\\
BDT score&$<0.08$&$\geq0.08$&$<0.08$ &$<-0.15$ &$<0.08$&$<0.1$&$<0.08$\\
\bottomrule
\end{tabular}
\label{tab:CR}
\end{table}

To classify the events as signal or background, the
AdaBoost BDT
algorithm~\cite{FREUND1997119, breiman2017classification} is used as implemented
in the scikit-learn package~\cite{JMLR:v12:pedregosa11a}. The training and optimization
of the BDT are performed in two steps. The first step is the optimization of
the BDT hyperparameters and the second step is the optimization of the training
through the selection of the variables used for the training.
 
To optimize the hyperparameters (maximum tree depth, maximum features per split,
minimum samples per leaf, minimum samples per split, number of estimators, and
learning rate) the set of 34 kinematic and topological variables
listed in Table~\ref{tab:inputvars} were used. These variables are chosen as they
are expected to provide good separation between the background and signal topologies.
To avoid biasing the search by selecting a specific mass, the
simulated signal samples described in Section~\ref{sec:datasim} are
combined with equal weight for \Mtaup
set to \SIlist{800; 900; 1000}{\GeV} to train the BDT.
All of the background samples are considered in the training, including all fake leptons.
A scan over the hyperparameters
is performed using the 5-fold cross-validation procedure
to train the BDT, where the simulated data samples are split into five
equal randomized samples, with four being used in the training and the fifth
used as a testing sample. The training is performed five times so that each
of the five samples is used as a testing sample. The set of hyperparameters with
the highest receiver operating characteristic (ROC) score is selected.

To select the optimal set of training variables, a BDT is trained in
each analysis region defined by lepton multiplicity using the full set of 34 variables.
The ranking of the variables is evaluated using the procedure provided by the
scikit-learn package.
The lowest-ranked training variable is then
removed and the BDT is retrained. This procedure is repeated until
the ROC score decreases by more than 1\%. This leads to each SR having its
own unique set of training variables, which are listed in Table~\ref{tab:vars}.
 
\begin{figure}[htbp]
\centering
\subfloat[]{
\includegraphics[scale = 0.26]{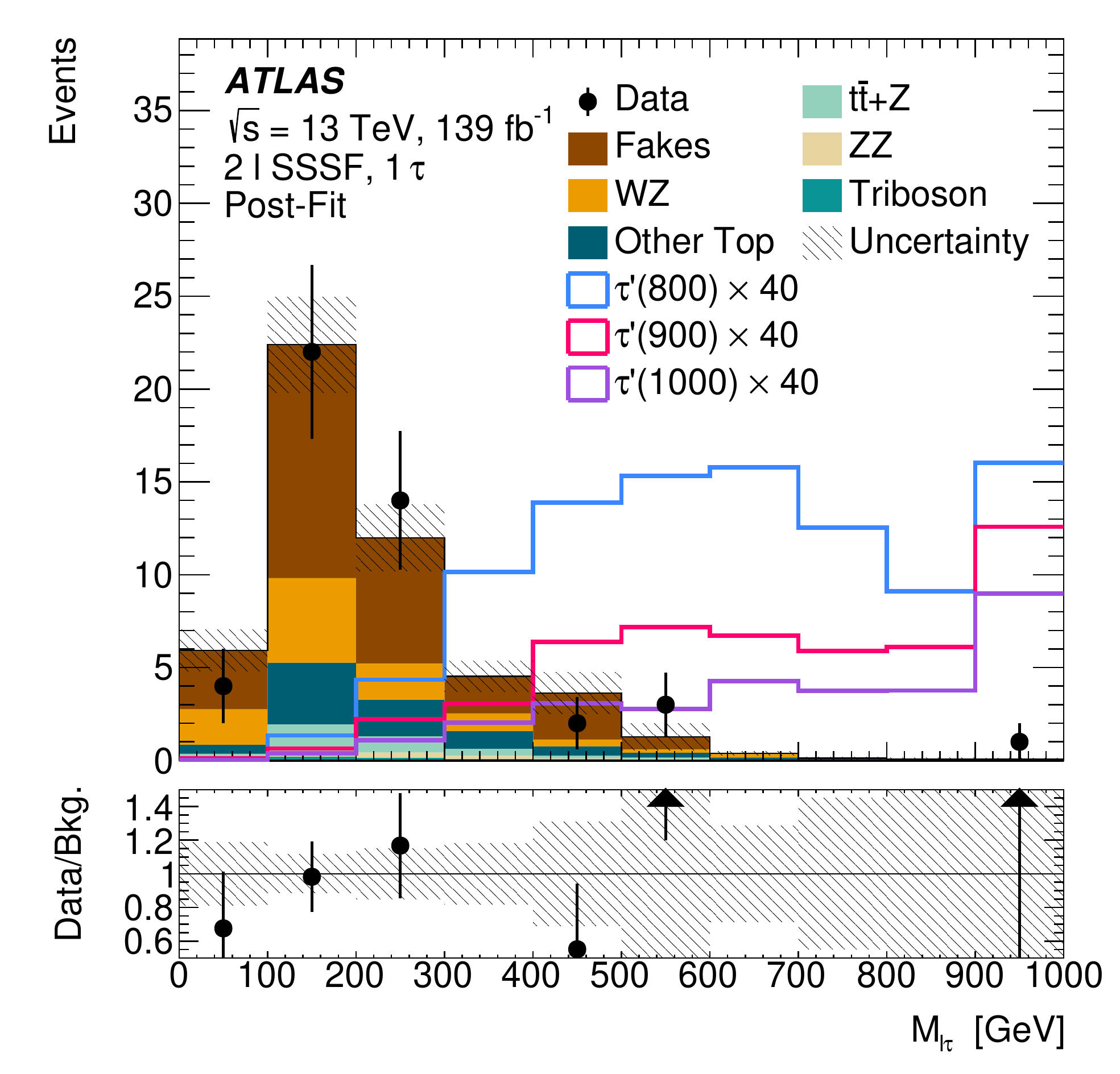}
\label{fig:invariablesa}
}
\subfloat[]{
\includegraphics[scale = 0.26]{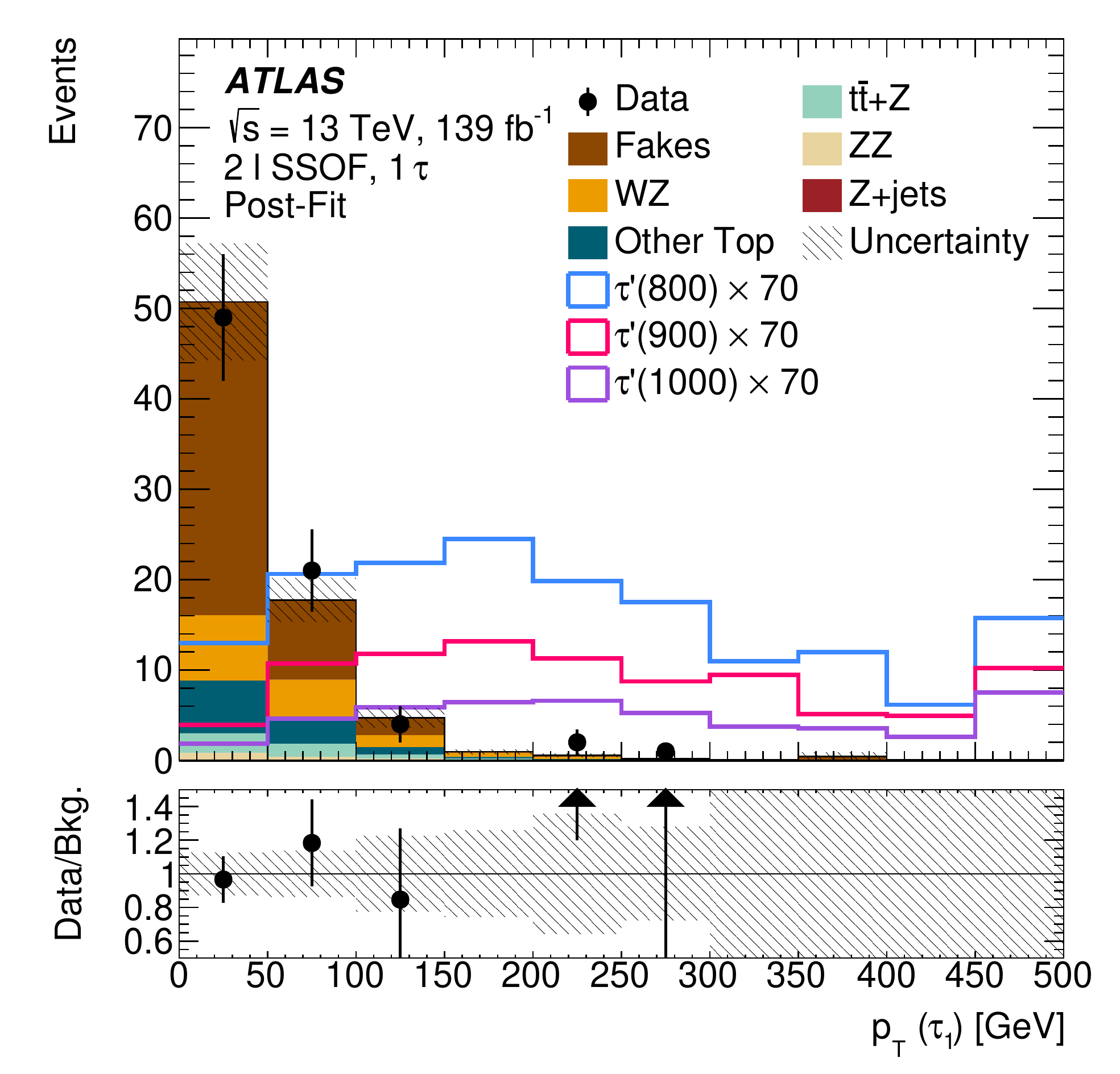}
\label{fig:invariablesb}
}
\subfloat[]{
\includegraphics[scale = 0.26]{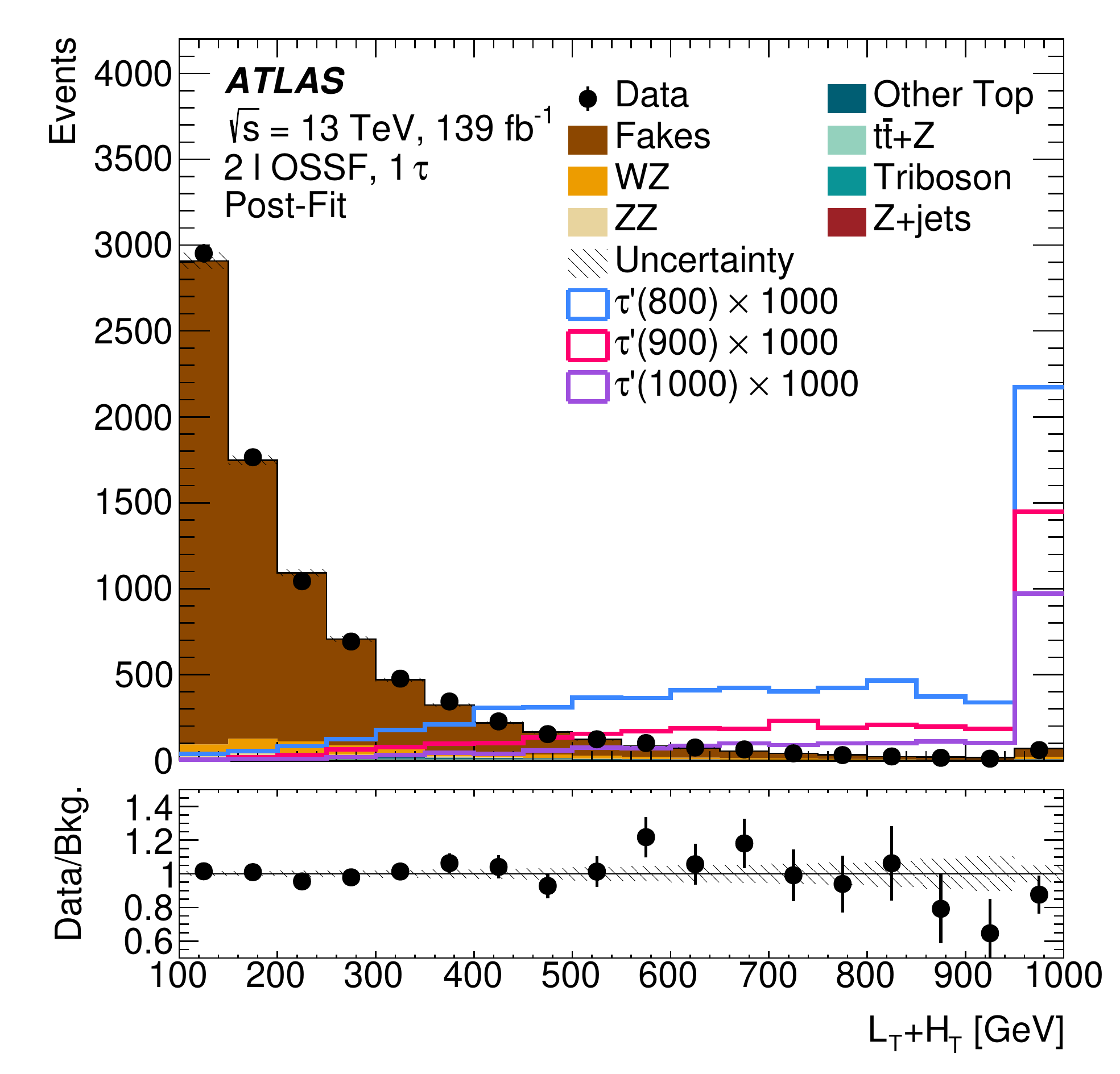}
\label{fig:invariablesc}
}
\hfill
\subfloat[]{
\includegraphics[scale = 0.26]{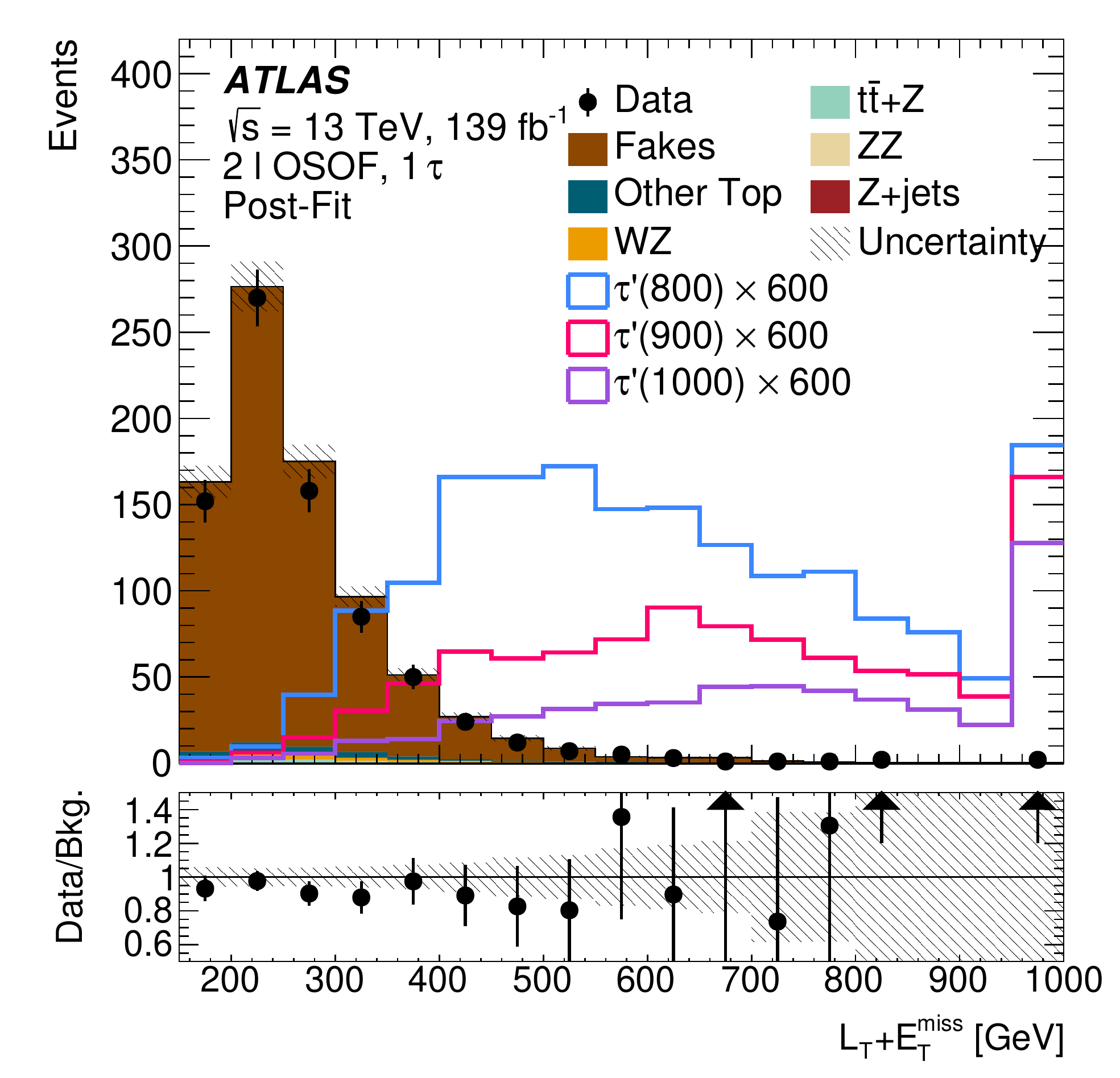}
\label{fig:invariablesd}
}
\subfloat[]{
\includegraphics[scale = 0.26]{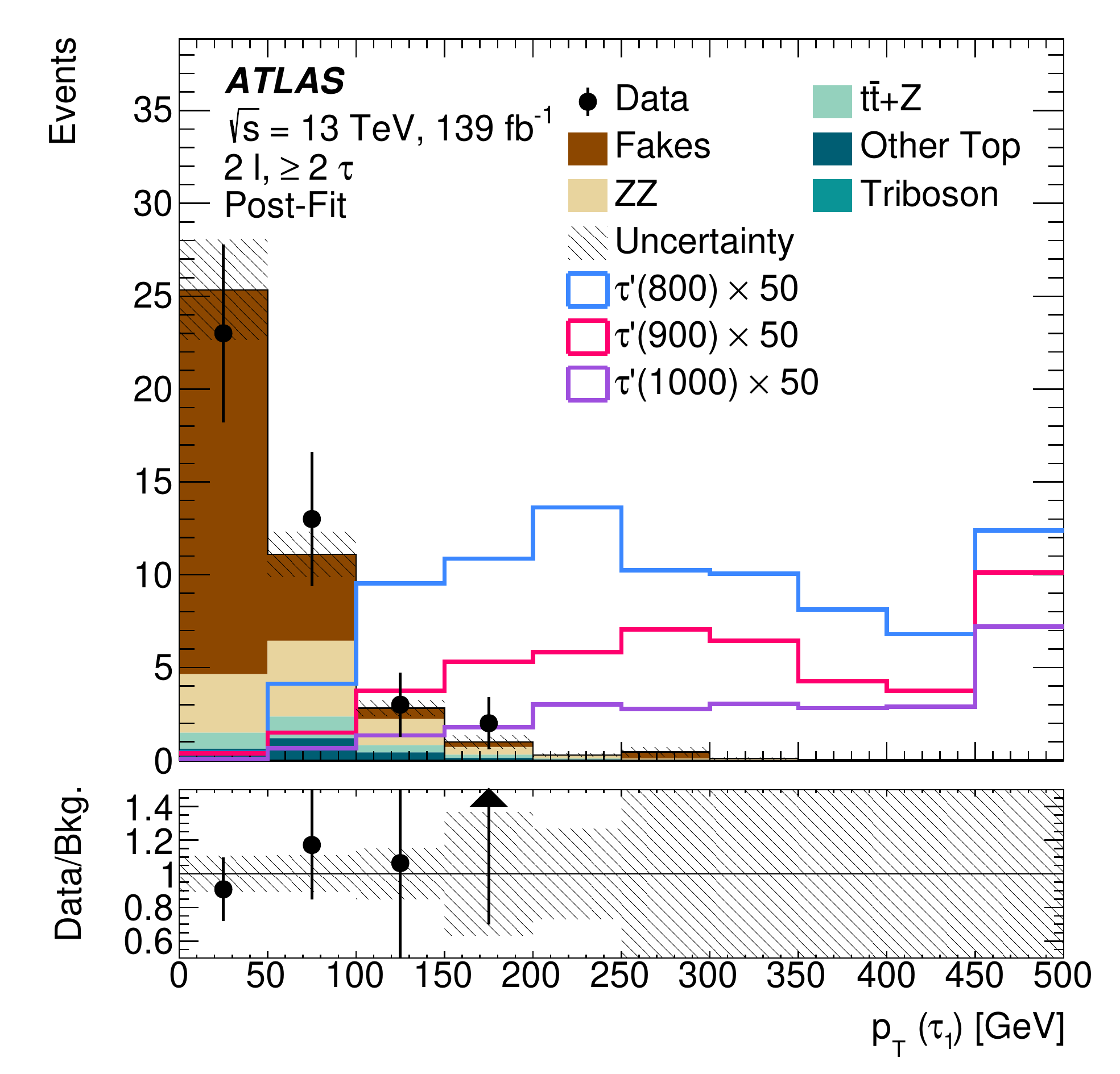}
\label{fig:invariablese}
}
\subfloat[]{
\includegraphics[scale = 0.26]{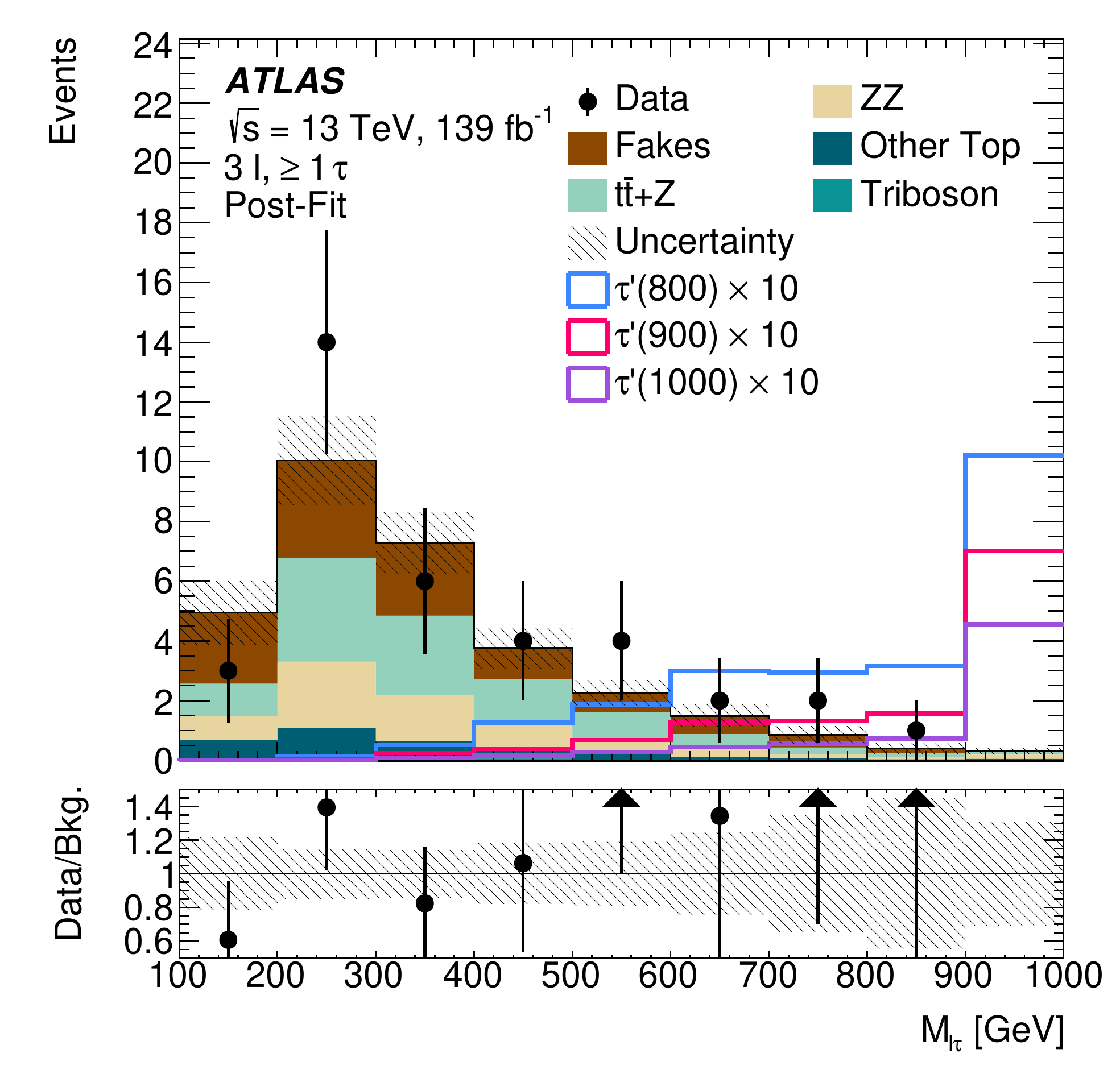}
\label{fig:invariablesf}
}
\hfill
\subfloat[]{
\includegraphics[scale = 0.26]{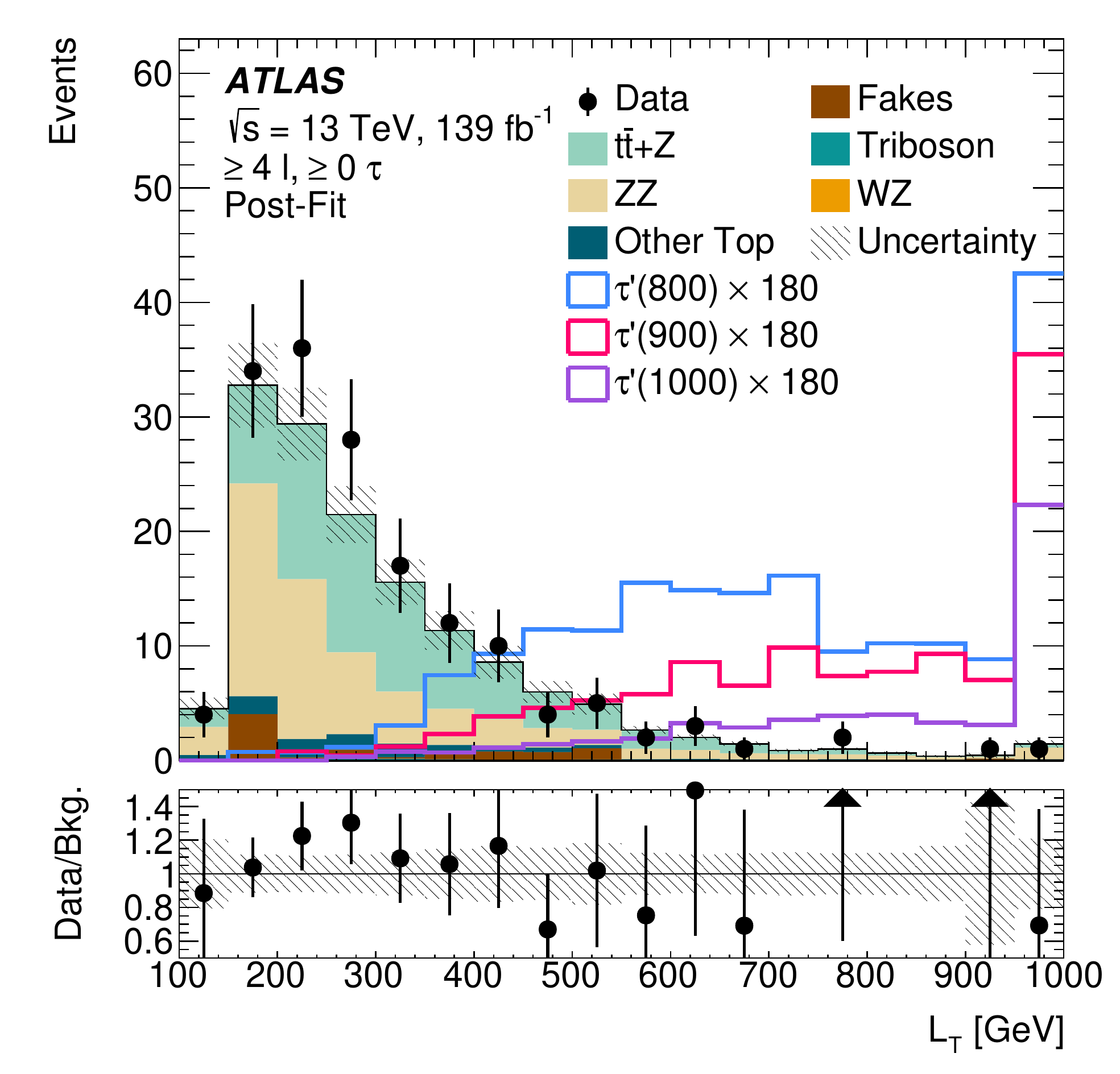}
\label{fig:invariablesg}
}
\caption{A comparison of the total background and signal distributions for
$M_{\tau'}= 800$, 900 and \SI{1000}{\GeV} in variables that are highly ranked in the BDT training. The signal distribution is scaled by the value indicated in the legend. The background prediction is taken after performing the fit described in Section~\ref{sec:results}, while the signal prediction is taken before the fit. The hatched band represents the combined statistical and systematic uncertainties. The last bin contains overflow events. The arrows in the ratio plot are for points that are outside the range.
\protect\subref{fig:invariablesa} $M_{\ell\tau}$ for the 2 $\ell$ SSSF, 1 $\tau$ training region \protect\subref{fig:invariablesb} $\pt(\tau_{1})$ for the 2 $\ell$ SSOF, 1 $\tau$ training region \protect\subref{fig:invariablesc} $L_\text{T}+H_\text{T}$ for the 2 $\ell$ OSSF, 1 $\tau$ training region \protect\subref{fig:invariablesd} $L_\text{T}+\MET$ for the 2 $\ell$ OSOF, 1 $\tau$ training region \protect\subref{fig:invariablese} $\pt(\tau_{1})$ for the 2 $\ell$, $\geq$ 2 $\tau$ training region \protect\subref{fig:invariablesf} $M_{\ell\tau}$ for the 3 $\ell$, $\geq$ 1 $\tau$ training region \protect\subref{fig:invariablesg} $L_\text{T}$ for the $\geq$ 4 $\ell$, $\geq$ 0 $\tau$ training region.
}
\label{fig:invariables}
\end{figure}

The final training variables are individually compared with data
to confirm that they are well modelled. To avoid any bias in the analysis, only events
with a BDT score not satisfying the SR criteria (Table~\ref{tab:SR}) are used in
the comparison, since they are background dominated.
Through the use of a $\chi^2$ test, a probability of $>5\%$ is found for agreement between data and simulation
for all variables used in the BDT training.
Figure~\ref{fig:invariables} shows the distributions of some highly ranked variables in each of the BDT training regions.

 
\begin{table}[htbp]
\caption{List of the input variables used to train the BDT. The final set is reduced by assessing the impact of removing the lowest-ranked variables on the ROC score for each training region independently.}
\begin{center}
\resizebox{\textwidth}{!}{
\begin{tabular}{ll}
\toprule
Variable& Description\\
\midrule
\MET& The missing transverse momentum in the event\\
$\mathcal{S}(\MET)$& The missing transverse momentum's significance in the event\\
$\LT$& The scalar sum of light lepton $\pT$ in the event\\
$\LT + \MET$& The scalar sum of light lepton $\pT$ and the missing transverse momentum in the event\\
$\LT + \pT(\tau)$& The scalar sum of light lepton $\pT$ and $\tau$-lepton $\pT$ in the event\\
$\pT(\ell_{1})$& The leading light lepton's $\pT$ in the event\\
$\pT(\ell_{2})$& The sub-leading light lepton's $\pT$ in the event\\
$\pT(j_{1})$& The leading jet's $\pT$ in the event\\
$\pT(\tau_{1})$& The leading $\tau$-lepton's $\pT$ in the event\\
$N_{j}$& The number of jets in the event\\
$N_{b}$& The number of $b$-jets in the event\\
$H_\text{T}$& The scalar sum of jet $\pT$ in the event\\
$\LT + H_\text{T}$& The scalar sum of light lepton $\pT$ and jet $\pT$ in the event\\
$M_{\ell\ell}$& The invariant mass of all light leptons in the event\\
$M_{\ell\tau}$& The invariant mass of all light leptons and $\tau$-leptons in the event\\
$M_{\ell j}$& The invariant mass of all light leptons and jets in the event\\
$M_{jj}$& The invariant mass of all jets in the event\\
$M_{j\tau}$& The invariant mass of all jets and $\tau$-leptons in the event\\
$\MT$& The transverse mass of the leading light lepton and \MET in the event\\
$M_{\text{OSSF}}$& The invariant mass of the opposite-sign same-flavour light-lepton pair closest to the $Z$ mass in the event\\
$\Delta\phi(j_{1}\MET)$& $\Delta\phi$ between the leading $\pT$ jet in the event and $\MET$\\
$\Delta\phi(\ell_{1}\MET)$& $\Delta\phi$ between the leading $\pT$ light lepton in the event and $\MET$\\
$\Delta\phi(\ell_{1}\ell_{2})$& $\Delta\phi$ between the leading and sub-leading $\pT$ light leptons in the event\\
$\Delta\phi(\ell_{1}j_{1})$& $\Delta\phi$ between the leading $\pT$ light lepton and jet in the event\\
$\Delta\phi(\tau_{1}\MET)$& $\Delta\phi$ between the leading $\pT$ $\tau$-lepton in the event and $\MET$\\
$\Delta\phi(\ell_{1}\tau_{1})$& $\Delta\phi$ between the leading $\pT$ light lepton and $\tau$-lepton in the event\\
$\Delta\phi(j_{1}\tau_{1})$& $\Delta\phi$ between the leading $\pT$ jet and $\tau$-lepton in the event\\
$\Delta R(j_{1}\MET)$& $\Delta R$ between the leading $\pT$ jet in the event and $\MET$\\
$\Delta R(\ell_{1}\MET)$& $\Delta R$ between the leading $\pT$ light lepton in the event and $\MET$\\
$\Delta R(\ell_{1}\ell_{2})$& $\Delta R$ between the leading and sub-leading $\pT$ light leptons in the event\\
$\Delta R(\ell_{1}j_{1})$& $\Delta R$ between the leading $\pT$ light lepton and jet in the event\\
$\Delta R(\tau_{1}\MET)$& $\Delta R$ between the leading $\pT$ $\tau$-lepton in the event and $\MET$\\
$\Delta R(\ell_{1}\tau_{1})$& $\Delta R$ between the leading $\pT$ light lepton and $\tau$-lepton in the event\\
$\Delta R(j_{1}\tau_{1})$& $\Delta R$ between the leading $\pT$ jet and $\tau$-lepton in the event\\
\bottomrule
\end{tabular}
}
\end{center}
\label{tab:inputvars}
\end{table}


\begin{table}[htbp]
\caption{The ranking of the variables used as inputs to the BDT algorithm for each of the training regions. Variables are ranked relative to each other by counting the instances a particular variable is used by the BDT nodes in defining the signal--background separation. Variables that do not have a ranking are not included in the corresponding training region. The final variables in each training region are selected using an optimization procedure.}
\centering
\resizebox{\textwidth}{!}{
\begin{tabular}{l|ccccccc}
\toprule
Variable &2$\ell$ SSSF, 1$\tau$&2$\ell$ SSOF, 1$\tau$&2$\ell$ OSSF, 1$\tau$&2$\ell$ OSOF, 1$\tau$&2$\ell$, $\geq$2$\tau$&3$\ell$, $\geq$1$\tau$&4$\ell$, $\geq$0$\tau$\\
\midrule
$\pT(\tau_{1})$	&\phantom{0}1&\phantom{0}1&\phantom{0}1&\phantom{0}1&\phantom{0}2&\phantom{0}2&\\
$M_{\ell\tau}$&\phantom{0}2&\phantom{0}2&\phantom{0}5&\phantom{0}3&\phantom{0}1&\phantom{0}1&\\
$\LT + \MET$&\phantom{0}3&\phantom{0}3&\phantom{0}2&\phantom{0}2&23&\phantom{0}4&\phantom{0}1\\
$\MET$&\phantom{0}4&\phantom{0}7&\phantom{0}4&21&\phantom{0}5&\phantom{0}8&\phantom{0}5\\
$\Delta\phi(\tau_{1}\MET)$&\phantom{0}5&\phantom{0}6&\phantom{0}6&13&\phantom{0}3&\phantom{0}3&\\
$\Delta R(\ell_{1}\ell_{2})$&\phantom{0}6&24&\phantom{0}7&\phantom{0}7&15&&17\\
$M_{jj}$&\phantom{0}7&21&24&15&\phantom{0}1&12&19\\
$M_{\ell j}$&\phantom{0}8&11&26&11&27&14&\phantom{0}2\\
$\Delta\phi(\ell_{1}\MET)$&\phantom{0}9&16&20&\phantom{0}8&20&10&15\\
$\Delta R(\ell_{1}\tau_{1})$&10&\phantom{0}8&12&\phantom{0}6&16&15&\\
$\Delta R(j_{1}\tau_{1})$&11&\phantom{0}9&17&25&25&23&\\
$\Delta R(\ell_{1}\MET)$&12&29&11&19&17&11&10\\
$\Delta\phi(\ell_{1}\ell_{2})$&13&13&18&16&28&13&\phantom{0}9\\
$\Delta R(\tau_{1}\MET)$&14&27&\phantom{0}9&\phantom{0}5&12&\phantom{0}9&\\
$\pT(j_{1})$&15&19&10&12&22&19&11\\
$\MT$&16&23&16&18&\phantom{0}8&17&\phantom{0}7\\
$\Delta\phi(j_{1}\tau_{1})$&17&20&27&29&24&&\\
$M_{\ell\ell}$&18&10&25&20&10&22&\phantom{0}4\\
$\pT(\ell_{1})$&19&\phantom{0}4&&&30&\phantom{0}5&16\\
$\mathcal{S}(\MET)$&20&\phantom{0}5&14&24&\phantom{0}9&24&\phantom{0}8\\
$N_{j}$&21&14&28&23&26&&22\\
$\LT + \pT(\tau)$&22&22&&26&&&\\
$\pT(\ell_{2})$&23&15&&&18&&\\
$\Delta R(j_{1}\MET)$&24&18&23&10&31&&21\\
$\Delta\phi(\ell_{1}j_{1})$&25&17&13&17&13&25&13\\
$N_{b}$&26&26&21&22&29&20&14\\
$\LT$&27&32&&&32&&\phantom{0}3\\
$M_{j\tau}$&28&31&15&\phantom{0}9&\phantom{0}6&18&\\
$\Delta R(\ell_{1}j_{1})$&29&&\phantom{0}8&\phantom{0}4&11&&18\\
$\LT + \HT$&&12&\phantom{0}3&14&&&\\
$M_{\text{OSSF}}$&&&22&&\phantom{0}7&\phantom{0}6&12\\
$\Delta\phi(\ell_{1}\tau_{1})$&&25&19&&19&16&\\
$\Delta\phi(j_{1}\MET)$&&&&27&21&&\phantom{0}6\\
$\HT$&&28&&28&33&21&20\\
\bottomrule
\end{tabular}
}
\label{tab:vars}
\end{table}


\section{Background estimation}
\label{sec:bkg}

There are two basic categories of backgrounds to the signal. One category,
the irreducible backgrounds, is defined by those processes
that yield the same final state as the signal. The other category,
the reducible backgrounds,
is defined by those that mimic the final state because of misidentified leptons
or non-prompt leptons as well as misidentified lepton charge.
The irreducible backgrounds are estimated from the simulated samples
discussed in Section~\ref{sec:datasim}.
 
Similarly to Ref.~\cite{HDBS-2019-06}
the simulated background in the \PW{}\PZ CR is found not to agree with
data when examined as a function of the number of jets in
the event. Since inverting the BDT score criteria for this CR yielded a similar
mismodelling, the data in this region are used to calculate a
scale factor to correct the MC simulation to the data in the \PW{}\PZ CR, which then agreed
with data.
 
Charge misidentification for electrons arises from photon conversions or bremsstrahlung,
and its rate is challenging to describe through detector simulation. Therefore, scale factors are derived
and applied to simulated background events to match the charge misidentification probabilities
observed in data. The scale factors are derived using a $Z\to e^+e^-$ data sample and are
parameterized as a function of $\pT$ and $\eta$ \cite{EGAM-2018-01}.
 
The reducible backgrounds from misidentified leptons (electron, muons and \tauhad) are estimated from
data by using the fake-factor (FF) method to derive a transfer function
from a background-dominated region to the SR~\cite{EGAM-2019-01}.
The transfer function
is the ratio of events passing the tight lepton selection to events passing
a loose lepton selection in the background-dominated CR.
The FF is used to estimate the probability of an object being
either a misidentified lepton or a non-prompt lepton from an
in-flight decay. The FF is calculated as a function of the lepton
\pT.
 
Non-prompt or fake light leptons can originate from decays of bottom
or charm hadrons, pion or kaon decays, jets misidentified as electrons,
and electrons from photon conversions.
To calculate the transfer function for electrons, the background
CR requires exactly one electron with a loose selection and no other leptons,
$\MET<\SI{40}{\GeV}$, $\Njet\geq 2$, and no reconstructed \Pqb-jets.
The background CR for muons requires exactly one muon with
a loose selection and zero electrons and $\tau$-leptons, $\MET<\SI{40}{\GeV}$,
and at least two jets with the leading jet $\pT>\SI{35}{\GeV}$.
 
Since only \tauhad decays are selected, misidentified $\tau$-leptons
originate from jets
that can arise from several sources: light- and heavy-flavour jets,
gluon radiation, and jets from \pileup.
The SRs are dominated by light- and heavy-flavour jets misidentified as $\tau$-leptons.
Therefore,
FFs are calculated independently in dedicated CRs corresponding to light- and heavy-flavour jets.
The \Zjets FF CR uses a \Zjets sample, which is enriched in
light-flavour fake $\tau$-leptons with a requirement of
\Zmpmm decay and $|\mmumu-\mZ|<\SI{15}{\GeV}$. In addition, the sample is
required to have zero reconstructed \Pqb-jets and $\MET<\SI{60}{\GeV}$.
The \ttbar FF CR is a dilepton \ttbar sample, enriched in heavy-flavour fake $\tau$-leptons with a
requirement of exactly two light leptons that satisfy
$|\mll-\mZ|>\SI{10}{\GeV}$, exactly one \Pgt-lepton that satisfies the
loose criteria, and at least two jets with at least one satisfying the \Pqb-jet
requirements. The FF is calculated separately for 1- and 3-prong
$\tau$-leptons. The FFs calculated in the \Zjets and \ttbar CRs are combined by performing
a template fit of the BDT score in each training region. The fit uses MC \enquote{truth}-matched fake $\tau$-leptons to
compare the fake composition in the analysis region and the \Zjets and \ttbar CRs.
The modelling of the background originating from fake leptons is verified by performing closure tests in the regions where
the FFs are derived as well as by comparing the background estimates to data in the VRs.


\section{Systematic uncertainties}
\label{sec:syst}

The systematic uncertainties considered in this analysis come from instrumental and theoretical
sources, affecting both the overall event yield and the shape of the distribution. They are evaluated
by varying each source around its nominal value as described below.
 
The uncertainties in the theoretical production cross sections used
to simulate the background events are calculated following the same
approach as in Ref.~\cite{EXOT-2019-36}. These uncertainties are assigned
to every background process whose normalization is not determined by the fit.
These backgrounds come from \Zjets, \PW{}\PW, triple vector boson, and top-quark processes
(excluding $\ttbar+\PZ{}$).

Uncertainties from missing higher-order contributions are evaluated~\cite{Bothmann:2016nao} by applying seven
independent variations of the QCD factorization and renormalization scales by factors
of one-half and two in the matrix elements
after removing combinations that differ by a factor
of four.
The effect of the uncertainty in the strong coupling constant $\alphas(\mZ)=0.118$ as well as the
uncertainties in the nominal PDF set, used in the generation
of simulated events, is evaluated by following the PDF4LHC recommendations~\cite{Butterworth:2015oua}.
In addition, the modelling uncertainty due to the choice of generator for $\ttbar+\PZ$ production is evaluated by comparing samples from the nominal
generator $\MCatNLO\,{+}\,\PYTHIA[8]$ and the alternative generator \SHERPA[2.2.1].
This gives an uncertainty of up to 7.9\% in the $\ttbar+\PZ$ background estimate.
 
The uncertainty in the
full integrated luminosity as obtained from the LUCID-2 detector is 1.7\%~\cite{Avoni2018iuv} and is applied to the background and signal processes
that are normalized to the theoretical predictions.
Uncertainties associated with the \pileup reweighting procedure range from
0.2\% to 3.5\%~\cite{ATLAS-CONF-2014-018}.

Instrumental uncertainties are evaluated for each object that is considered in the analysis. For the selected leptons,
they originate from the reconstruction and identification
efficiency, the energy (momentum) scale and resolution, and the isolation
efficiency~\cite{EGAM-2018-01, MUON-2018-03, PERF-2015-10}.
The uncertainty related to trigger efficiencies is also included. For jets,
the uncertainties originate from the jet energy scale and resolution~\cite{PERF-2015-09,PERF-2016-04},
the matching of jets to the primary vertex~\cite{ATLAS-CONF-2014-018}, and the identification
of $b$-jets~\cite{FTAG-2018-01,FTAG-2019-07}. Furthermore, the impact of a possible
miscalibration of the soft-track component of the \MET is derived from
a data--MC comparison of the \pT balance of the hard and soft \MET
components~\cite{PERF-2016-07}.

Systematic uncertainties associated with the light-lepton FFs~\cite{EGAM-2019-01} account for the jet
composition differences between the region where the FFs are derived and the analysis regions where they are applied,
as well as the differences between the values of FFs calculated in regions with different selection requirements.
Additionally, the uncertainty arising from the modelling and limited sample sizes is taken into account. The uncertainties
are treated independently for each of the different sources of misidentified objects.
 
Uncertainties from several sources are evaluated for the FF derived for $\tau$-leptons. Most notably,
to account for the limited numbers of simulated events in the regions used to determine the composition of fake $\tau$-lepton samples,
the uncertainty of the fitted fractions from the template fit to the \Zjets and \ttbar FF CRs, is evaluated as the difference between
the two values 0\% and 100\%, for each SR.
Furthermore, a systematic uncertainty to account 
for the gluon- and \pileup-initiated fake $\tau$-lepton contribution is applied. 
To assess this uncertainty
a new CR is defined with a less restrictive RNN score requirement on the loose \tauhadvis in the
\Zjets FF CR, which increases this contribution from roughly 40\% to 60\% in the CR.
The systematic uncertainty in the estimated number of fake $\tau$-leptons is then taken as the
difference between applying the nominal FF in each SR and the FF calculated with the less restrictive RNN requirement.
The nominal FF is calculated by applying a minimum RNN requirement of 0.01,
while for the assessment of its systematic uncertainty a minimum value of 0.005 is used.
Since gluon- and \pileup-initiated fake $\tau$-leptons typically have low
$\pT$, the $\tau$-lepton FF systematic uncertainties are split into $<$\SI{40}{\GeV} and $>$\SI{40}{\GeV} regions and assessed independently.
The gluon- and \pileup-initiated
contribution is only taken for fake $\tau$-leptons with $\pT$ less than \SI{40}{\GeV}, and the large uncertainty of 28\% is
constrained in the Fake \tauhad CR (Table~\ref{tab:CR}), which is a highly populated region dominated by gluon- and \pileup-initiated fake $\tau$-leptons.
However, the gluon- and \pileup-initiated fake $\tau$-lepton fraction is less than 20\% in the SRs, where
the BDTs tend to select $\tau$-leptons with $\pT$ greater than \SI{40}{\GeV}, so this uncertainty does not have an impact on them.
Additionally, uncertainties are derived for the $\tau$-lepton FFs by accounting for the numbers of events in the \Zjets and \ttbar FF CRs.


\section{Results}
\label{sec:results}

In order to test for the presence of a VLL signal, the BDT score templates for signal and
background events are fitted to the data using a binned maximum-likelihood (ML) approach
in the RooFit and RooStats frameworks~\cite{RooFit,moneta2011roostats}.
The normalizations to data of the main background component modelled in the CR templates ($\PW{}\PZ$,
$\PZ{}\PZ$, $\ttbar+\PZ{}$), and SR VLL templates are allowed to vary without constraints. The other backgrounds'
normalizations are assigned Gaussian constraints based on their normalization uncertainties.
In addition,
the systematic uncertainties are included in the fit as nuisance parameters with
correlations across regions and processes taken into account.
To quantify the statistical significance of the fit and its resulting power to
reject the background-only hypothesis, a test statistic is constructed using the
profile likelihood ratio~\cite{Cowan:2010js}.
 
\begin{figure}[htbp]
\centering
\subfloat[]{
\includegraphics[scale = 0.4]{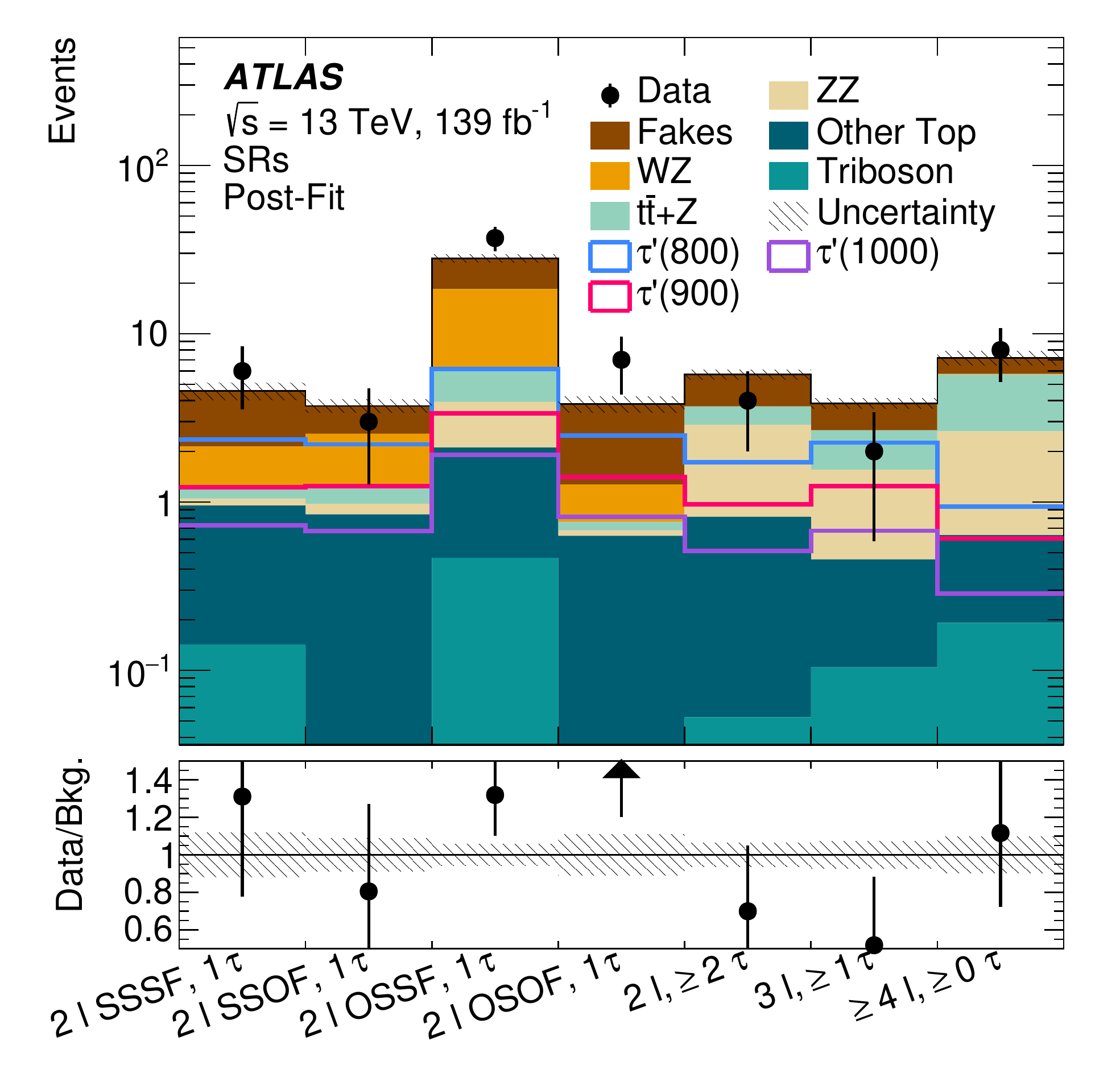}
\label{fig:summarySR}
}
\subfloat[]{
\includegraphics[scale = 0.4]{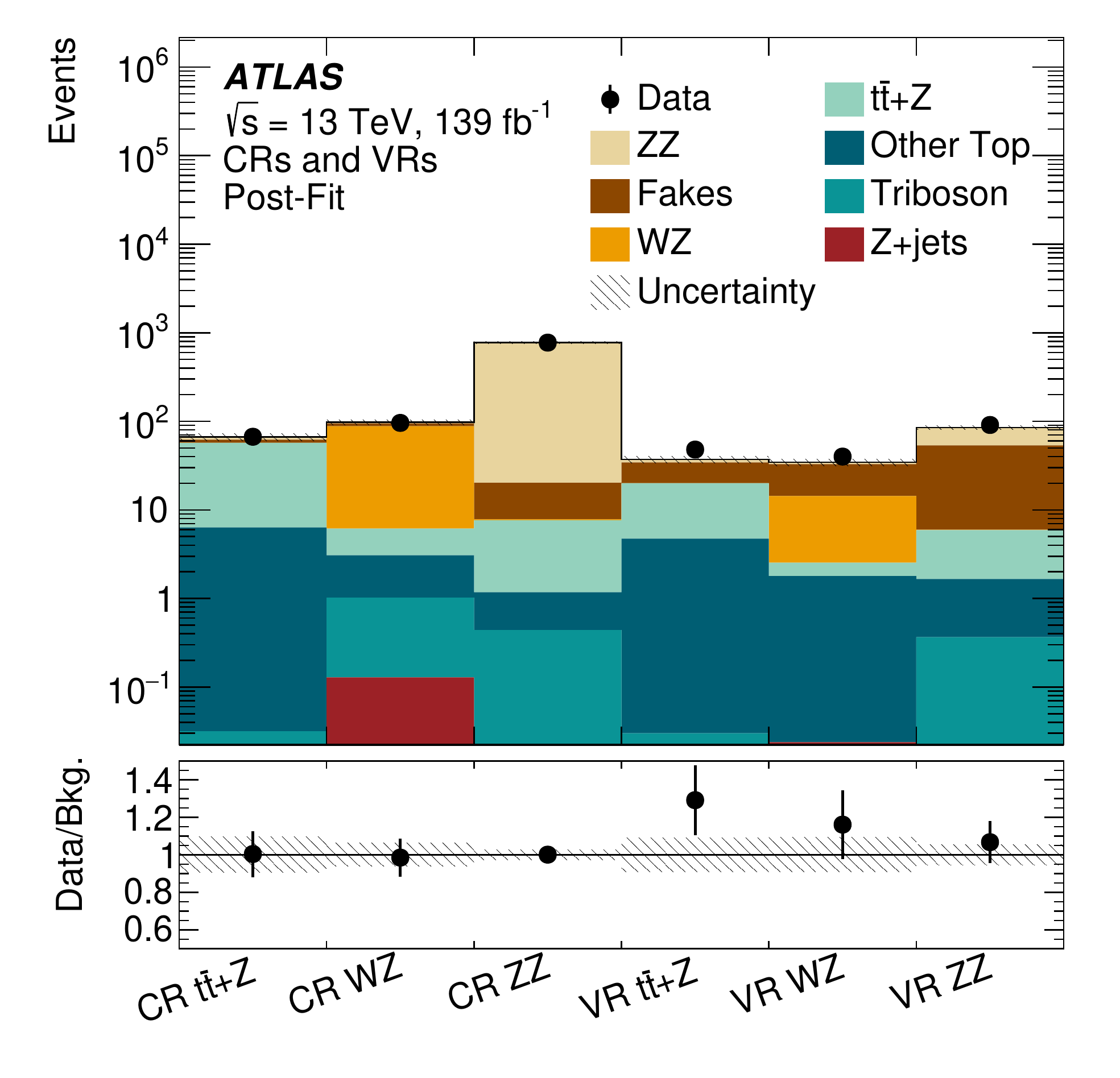}
\label{fig:summaryCR}
}
\caption{Summary of post-fit yields for data and background, and pre-fit yields for signal modelling in the \protect\subref{fig:summarySR} SRs and \protect\subref{fig:summaryCR} CRs and corresponding VRs. Uncertainty bands contain both the systematic and statistical uncertainties post-fit. Background normalization factors are also applied. The \enquote{Other Top} sample includes contributions from single top, \ttbar, \ttW, \tZ, \tWZ, \ttH, \ttWW, \tttt, and \Pqt{}\Pqt{}\Pqt. The arrow in the ratio plot is for the point that is outside the range.
\label{fig:summary}}
\end{figure}

\begin{figure}[htbp]
\centering
\subfloat[]{
\includegraphics[scale = 0.26]{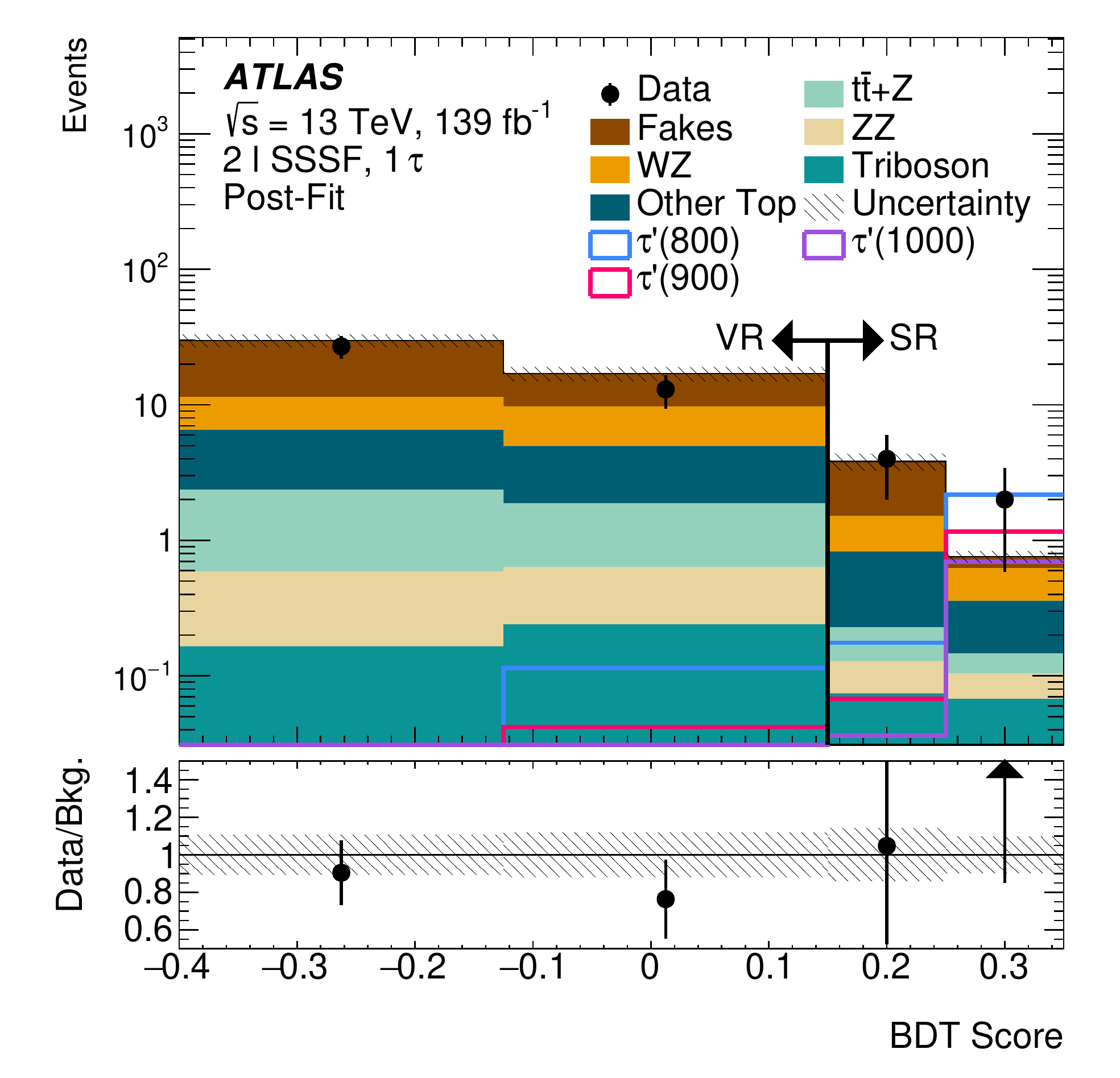}
\label{fig:BDTdistroa}
}
\subfloat[]{
\includegraphics[scale = 0.26]{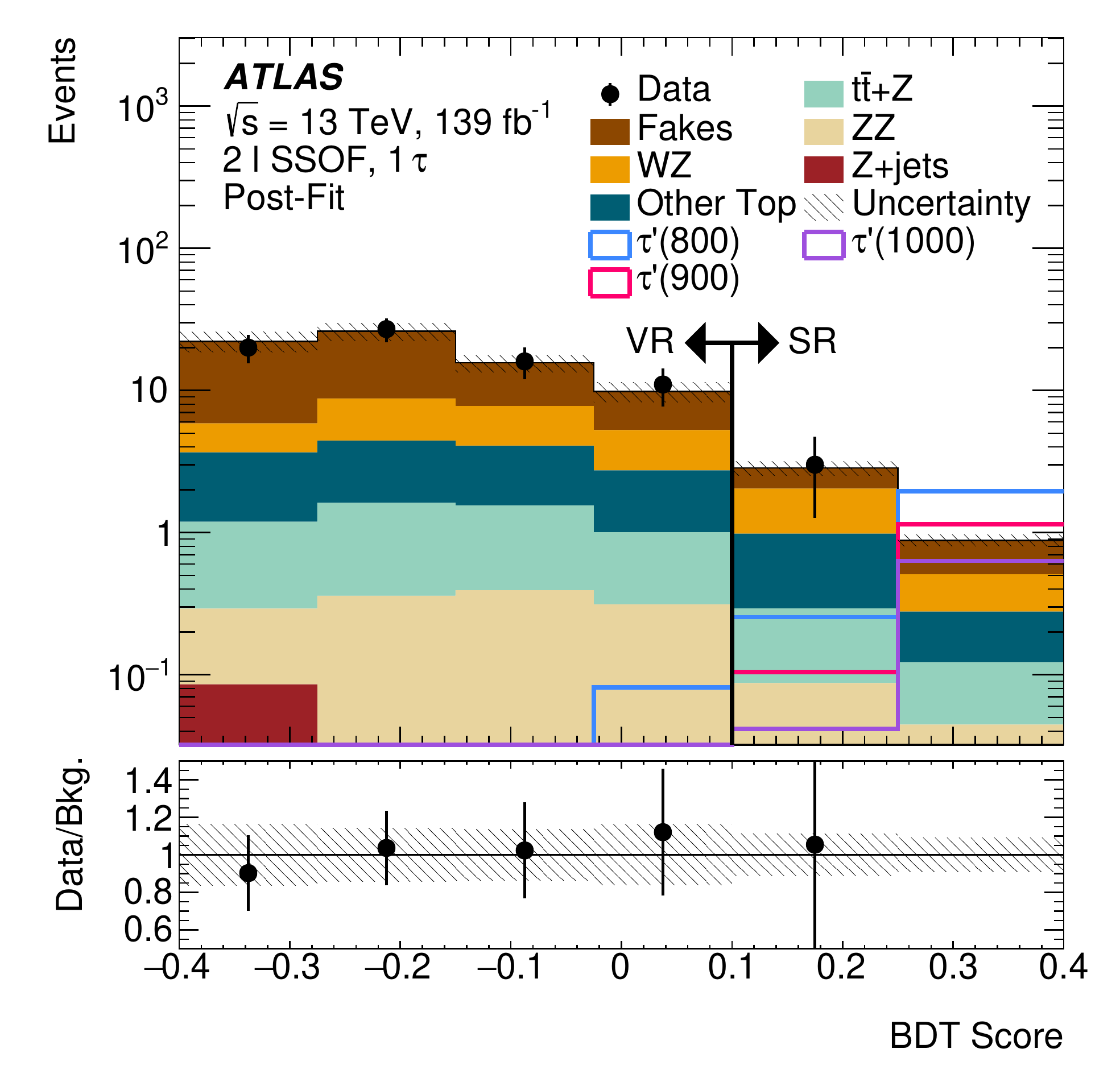}
\label{fig:BDTdistrob}
}
\subfloat[]{
\includegraphics[scale = 0.26]{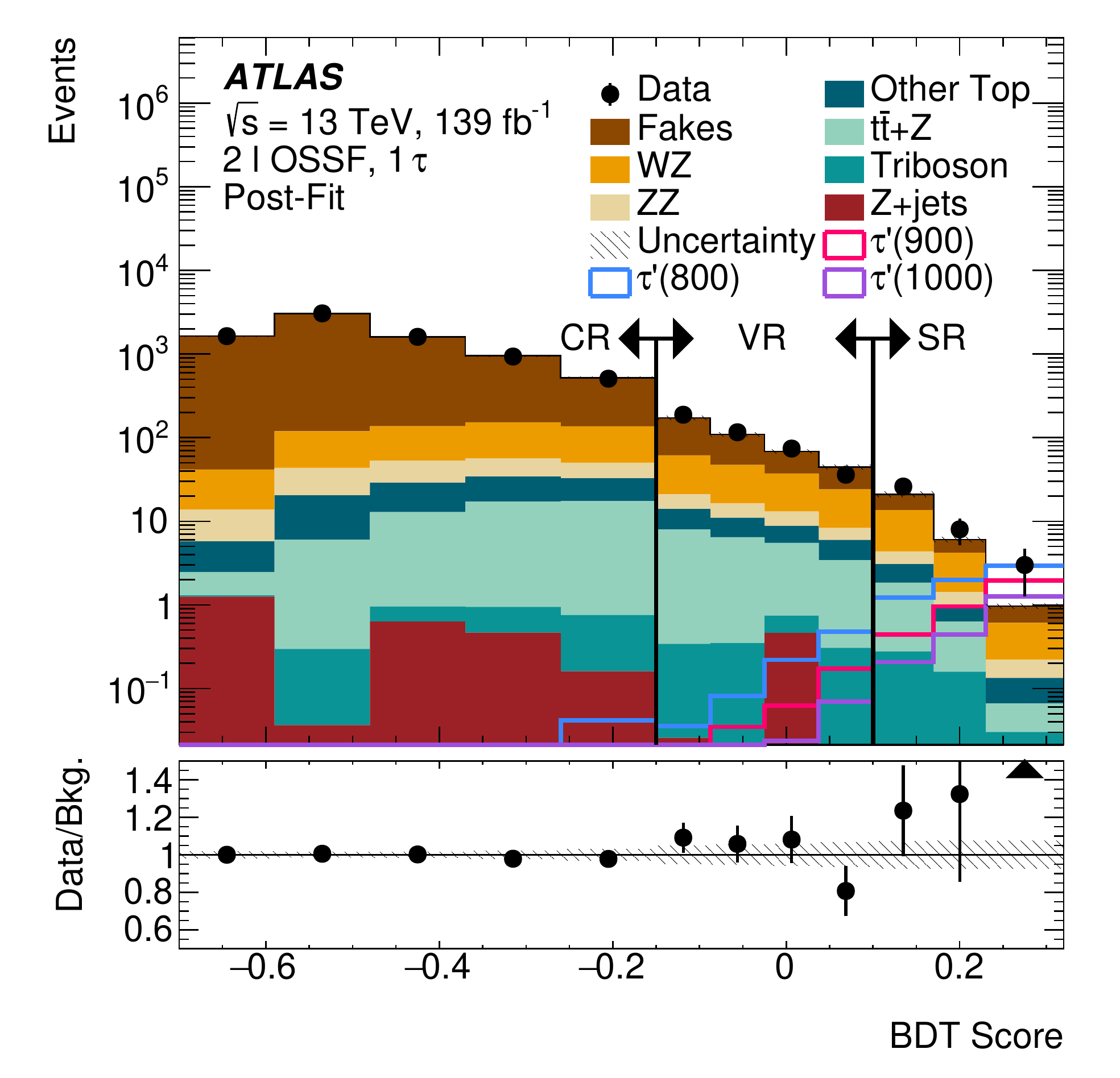}
\label{fig:BDTdistroc}
}
\hfill
\subfloat[]{
\includegraphics[scale = 0.26]{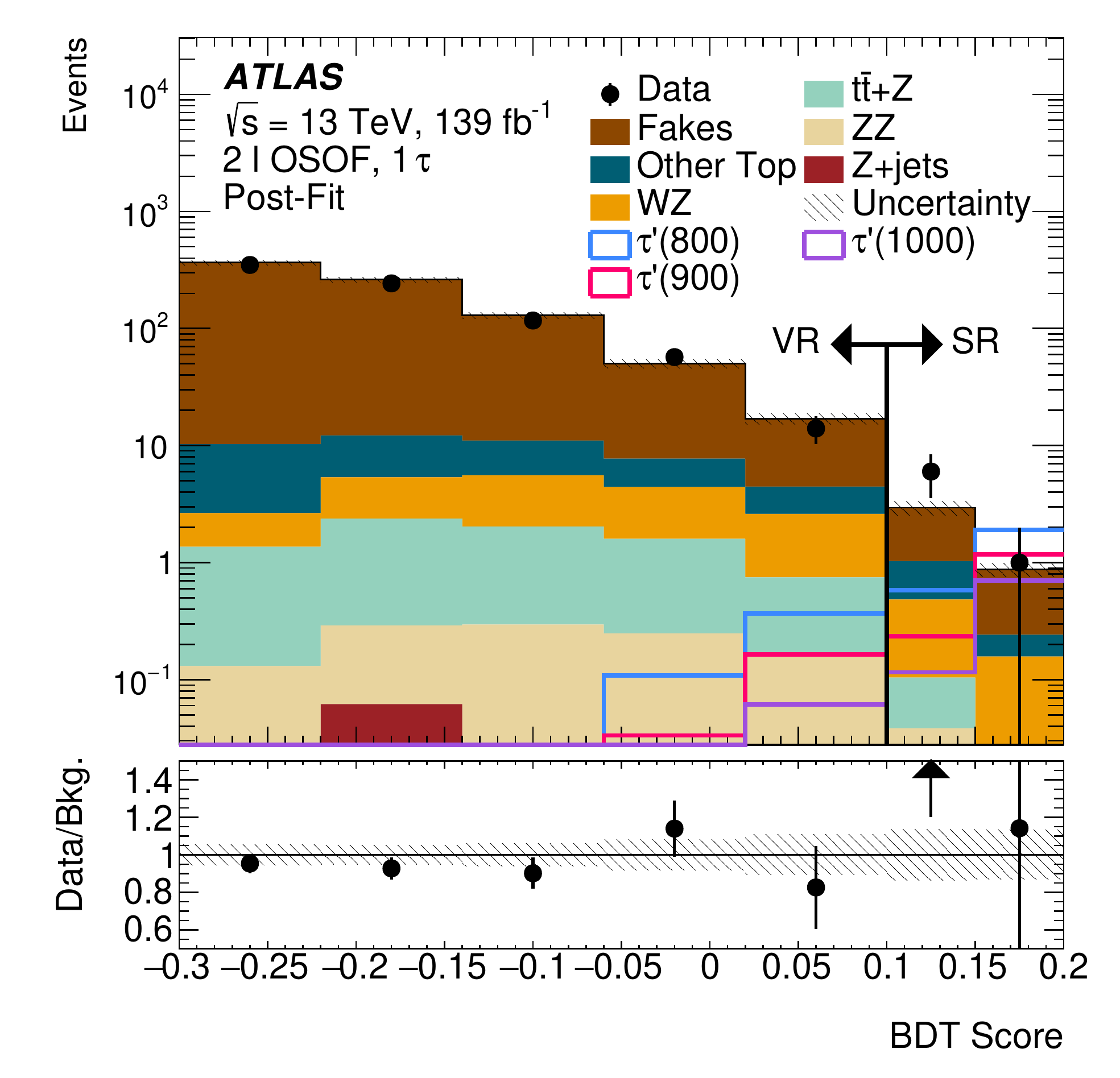}
\label{fig:BDTdistrod}
}
\subfloat[]{
\includegraphics[scale = 0.26]{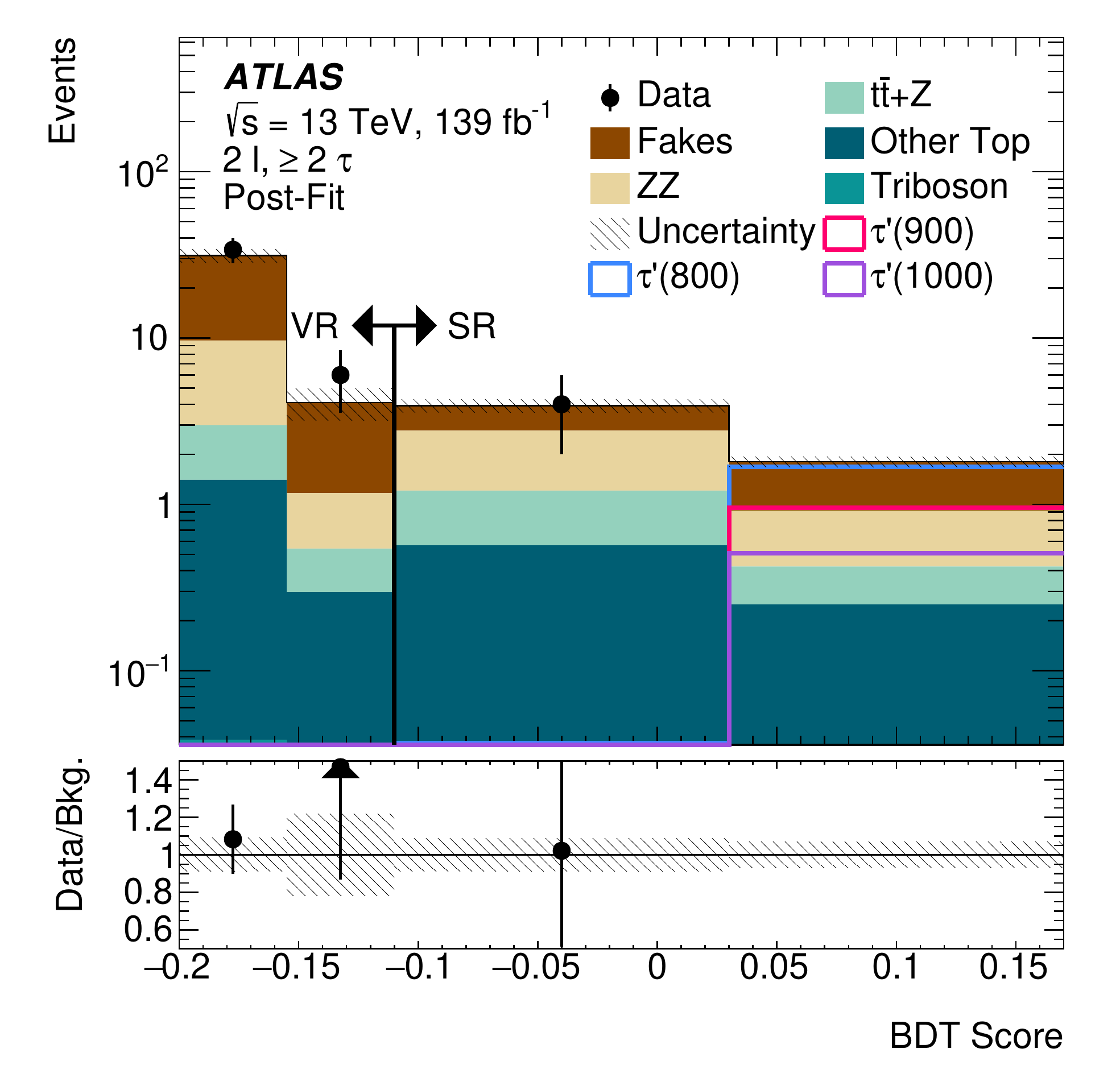}
\label{fig:BDTdistroe}
}
\subfloat[]{
\includegraphics[scale = 0.26]{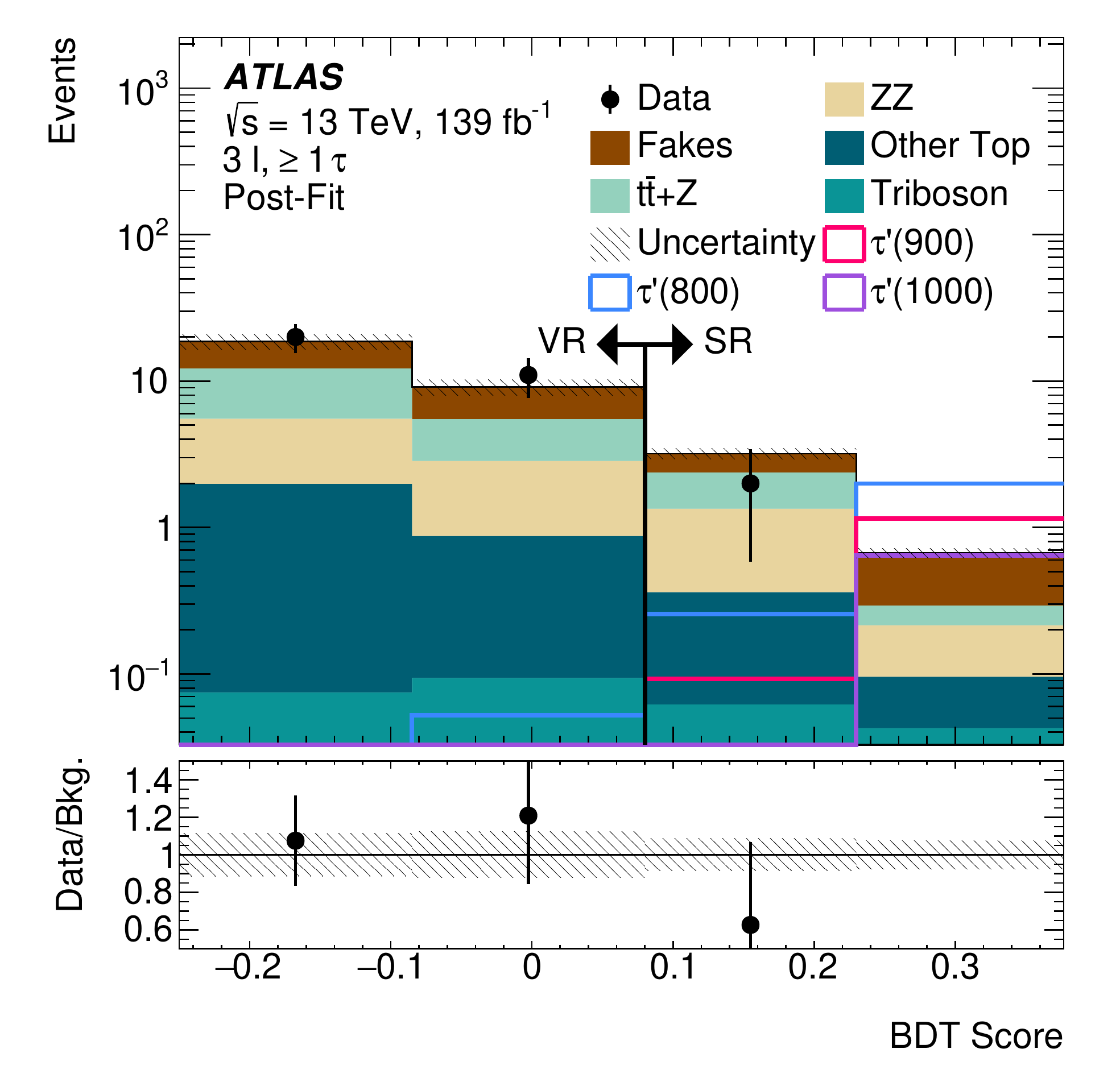}
\label{fig:BDTdistrof}
}
\hfill
\subfloat[]{
\includegraphics[scale = 0.26]{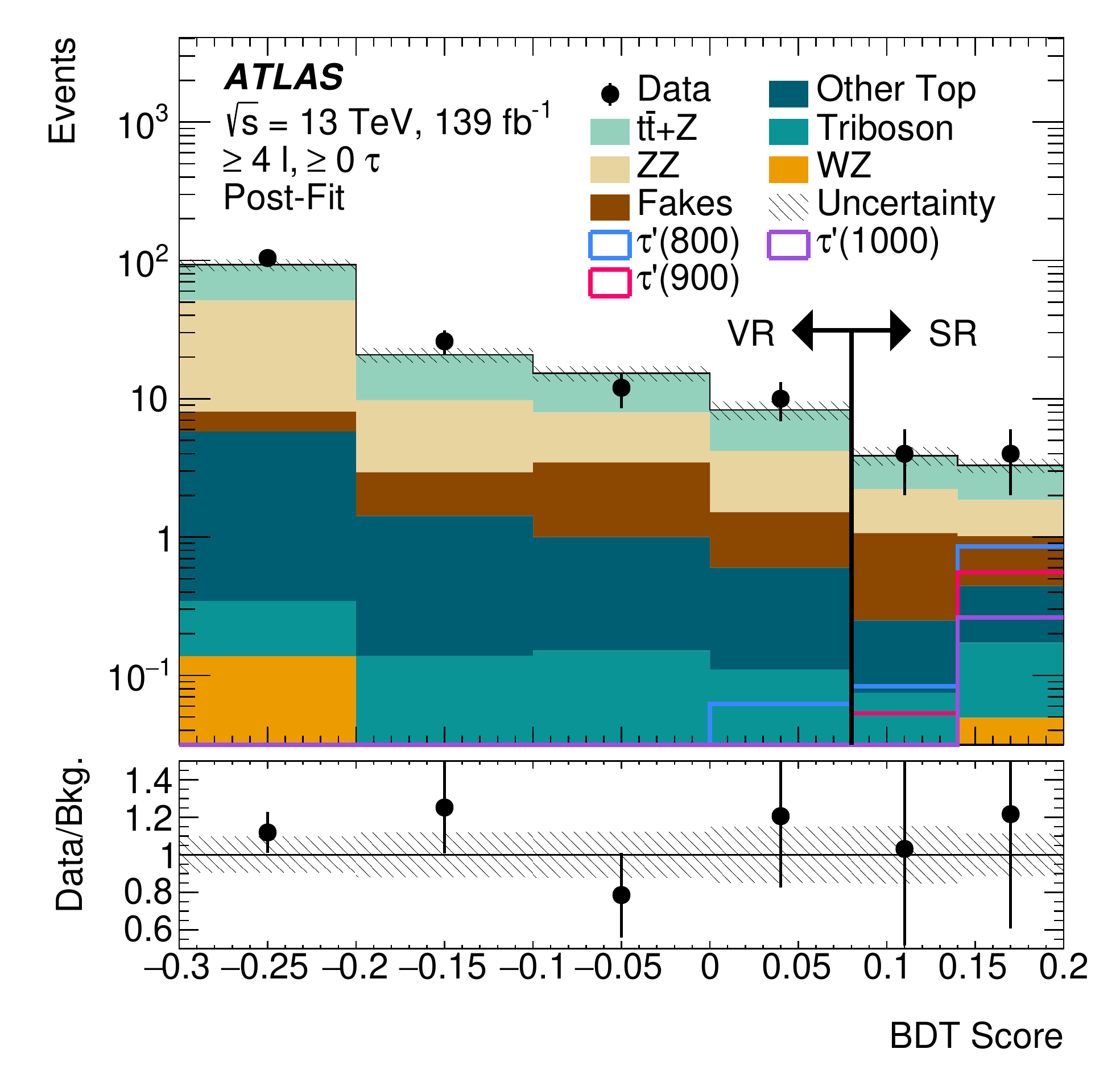}
\label{fig:BDTdistrog}
}
 
\caption{The post-fit BDT score distributions of data, background, and pre-fit signal modelling. The arrows indicate the point where there is a break between the regions and indicate the SRs used in the likelihood fit. The remaining distribution is treated as a VR and is not used in the fit (with the exception of \protect\subref{fig:BDTdistroc} 2~$\ell$ OSSF, where the low BDT score region is used as a CR and is included in the fit); however, fitted nuisance parameters are propagated to these regions. The uncertainty bands contain both the systematic and statistical components post-fit. The first bin contains underflow events and the last bin contains overflow events. The arrows in the ratio plot are for points that are outside the range.
\protect\subref{fig:BDTdistroa} 2~$\ell$ SSSF, 1~$\tau$; \protect\subref{fig:BDTdistrob} 2~$\ell$ SSOF, 1~$\tau$; \protect\subref{fig:BDTdistroc} 2~$\ell$ OSSF, 1~$\tau$; \protect\subref{fig:BDTdistrod} 2~$\ell$ OSOF, 1~$\tau$; \protect\subref{fig:BDTdistroe} 2~$\ell$, $\geq$~2~$\tau$; \protect\subref{fig:BDTdistrof} 3~$\ell$, $\geq$~1~$\tau$; \protect\subref{fig:BDTdistrog} $\geq$~4~$\ell$, $\geq$~0~$\tau$.}
\label{fig:BDTdistros}
\end{figure}

After performing the simultaneous fit of the seven SRs and four CRs
to the data,
the fitted normalization factors relative to the theoretical expectations
for the main backgrounds are $1.06\pm 0.14$ for $\PW{}\PZ$,
$1.02\pm 0.07$ for $\PZ{}\PZ$, and $1.18\pm 0.18$ for $\ttbar+\PZ{}$.
Figure~\ref{fig:summary} shows a comparison between the data and the
signal and background yields for the SRs, CRs, and corresponding VRs.
Figure~\ref{fig:BDTdistros} shows the templates and data versus BDT score for each of the
analysis regions after applying the selection requirements in Table~\ref{tab:SR} but before the SR's BDT
requirement is applied.
The regions with BDT score values less than the SR requirement are used as VRs.
In all SRs the number of observed events is compatible with the background hypothesis.
Tables~\ref{tab:yieldsSR} and \ref{tab:yieldsCR} show the total background and signal yields in all SRs and CRs after fitting to data.

\begin{table}[htbp]
\begin{center}
\caption{Total observed yields as computed by the fit for signal regions. The uncertainty contains both the systematic and statistical uncertainties. The prediction for each background sample is taken after a likelihood fit is performed to measure the VLL production cross section. Background normalization factors are also applied. The \enquote{Other Top} sample includes contributions from single top, \ttbar, \ttW, \tZ, \tWZ, \ttH, \ttWW, \tttt, and \Pqt{}\Pqt{}\Pqt. The prediction from the signal samples is taken before the likelihood fit is performed. The background contributions may not add up to equal the total background due to rounding.}
\resizebox{\textwidth}{!}{
\begin{tabular}{l|
S[table-format=1.3,table-figures-decimal=2,table-auto-round]@{\(\;\pm\;\)}S[table-format=1.1,table-figures-decimal=2,table-auto-round,table-number-alignment = left]
S[table-format=1.3,table-figures-decimal=2,table-auto-round]@{\(\;\pm\;\)}S[table-format=1.2,table-figures-decimal=2,table-auto-round,table-number-alignment = left]
S[table-format=2.3,table-figures-decimal=2,table-auto-round]@{\(\;\pm\;\)}S[table-format=1.2,table-figures-decimal=2,table-auto-round,table-number-alignment = left]
S[table-format=1.3,table-figures-decimal=2,table-auto-round]@{\(\;\pm\;\)}S[table-format=1.2,table-figures-decimal=2,table-auto-round,table-number-alignment = left]
S[table-format=1.3,table-figures-decimal=2,table-auto-round]@{\(\;\pm\;\)}S[table-format=1.2,table-figures-decimal=2,table-auto-round,table-number-alignment = left]
S[table-format=1.3,table-figures-decimal=2,table-auto-round]@{\(\;\pm\;\)}S[table-format=1.2,table-figures-decimal=2,table-auto-round,table-number-alignment = left]
S[table-format=1.3,table-figures-decimal=2,table-auto-round]@{\(\;\pm\;\)}S[table-format=1.2,table-figures-decimal=2,table-auto-round,table-number-alignment = left]}
\toprule
Signal Regions
& \multicolumn{2}{l}{\parbox{.15\textwidth}{\centering 2 $\ell$ SSSF, 1 $\tau$}}
& \multicolumn{2}{l}{\parbox{.15\textwidth}{\centering 2 $\ell$ SSOF, 1 $\tau$}}
& \multicolumn{2}{l}{\parbox{.15\textwidth}{\centering 2 $\ell$ OSSF, 1 $\tau$}}
& \multicolumn{2}{l}{\parbox{.15\textwidth}{\centering 2 $\ell$ OSOF, 1 $\tau$}}
& \multicolumn{2}{l}{\parbox{.15\textwidth}{\centering 2 $\ell$, $\geq$ 2 $\tau$}}
& \multicolumn{2}{l}{\parbox{.15\textwidth}{\centering 3 $\ell$, $\geq$ 1 $\tau$}}
& \multicolumn{2}{c}{\parbox{.15\textwidth}{\centering$\geq$ 4 $\ell$, $\geq$ 0 $\tau$}}
\\
\midrule
Observed Events   & \multicolumn{2}{l}{\phantom{0.00\;}\num{6}} & \multicolumn{2}{l}{\phantom{0.00\;}3} & \multicolumn{2}{l}{\phantom{0.00~\;}37} & \multicolumn{2}{l}{\phantom{0.00\;}7} & \multicolumn{2}{l}{\phantom{0.00\;}4} & \multicolumn{2}{l}{\phantom{0.00\;}2} & \multicolumn{2}{l}{\phantom{0.00\;}8} \\
\hline
Total Background  & 4.6 &0.6 & 3.7 &0.4 & 28.0 &1.8 & 3.8& 0.5 & 5.7& 0.4 & 3.9 &0.3 & 7.2 &0.7 \\
\hline
Other Top   & 0.81 & 0.19 & 0.85 & 0.24 & 1.6 & 0.4 & 0.63 & 0.19 & 0.76 & 0.09 & 0.35 & 0.07 & 0.44 & 0.13 \\
$t\bar{t}$+$Z$   &  0.14 & 0.04 & 0.28 & 0.07 & 2.1 & 0.4 & 0.08 & 0.05 & 0.82 & 0.16 & 1.11 & 0.2 & 3.1 & 0.6 \\
$ZZ$   & 0.091 & 0.023 & 0.132 & 0.016 & 1.83 & 0.14 & 0.048 & 0.013 & 2.06 & 0.28 & 1.1 & 0.12& 2.0 & 0.25 \\
$WZ$   &0.95 & 0.16 & 1.28 & 0.19 & 12.4 & 1.6 & 0.51 & 0.09 & \multicolumn{2}{l}{\parbox{.1\textwidth}{\centering < 0.01}}  &   \multicolumn{2}{l}{\parbox{.1\textwidth}{\centering < 0.01}} & 0.055 & 0.035 \\
Triboson   &  0.142 & 0.023& \multicolumn{2}{l}{\parbox{.1\textwidth}{\centering < 0.01}} & 0.47 & 0.04 &  \multicolumn{2}{l}{\parbox{.1\textwidth}{\centering < 0.01}}  & 0.053 & 0.005 & 0.104 & 0.01 & 0.192 & 0.022  \\
Fakes   & 2.4 & 0.5 & 1.18 & 0.17 & 9.6 & 1.5 & 2.5 & 0.4 & 2.03 & 0.17 & 1.2 & 0.13 & 1.39 & 0.34\\
\hline
$\tau'$(\SI{130}{\GeV})*   & 3.4 & 2.4 & 8. & 4. & 11. & 5. & 6. & 2.4 & 25. & 6. & 13. & 4. & 5.4 & 3.5  \\
 
$\tau'$(\SI{500}{\GeV})*   & 12.6 & 0.8 & 13.6 & 0.8 & 29.6 & 1.4 & 10.7 & 0.7 & 13.8 & 1. & 13.7 & 0.7 & 4.7 & 0.4 \\
 
$\tau'$(\SI{800}{\GeV})*   & 2.35 & 0.13 & 2.2 & 0.12 & 6.16 & 0.3 & 2.48 & 0.14 & 1.72 & 0.13 & 2.25 & 0.13 & 0.94 & 0.06 \\
 
$\tau'$(\SI{1000}{\GeV})*   & 0.73 & 0.04 & 0.67 & 0.04 & 1.9 & 0.09 & 0.82 & 0.04 & 0.51 & 0.04 & 0.68 & 0.04 & 0.286 & 0.02\\

\bottomrule
\multicolumn{8}{l}{*Pre-fit}
\end{tabular}
}
\label{tab:yieldsSR}
\end{center}
\end{table}


\begin{table}[htbp]
\begin{center}
\caption{Total observed yields as computed by the fit for control regions. The uncertainty contains both the systematic and statistical uncertainties. The prediction for each background sample is taken after a likelihood fit is performed to measure the VLL production cross section. Background normalization factors are also applied. The \enquote{Other Top} sample includes contributions from single top, \ttbar, \ttW, \tZ, \tWZ, \ttH, \ttWW, \tttt, and \Pqt{}\Pqt{}\Pqt. The prediction from the signal samples is taken before the likelihood fit is performed. The background contributions may not add up to equal the total background due to rounding.}
 
\begin{tabular}{l|cccc}
\toprule
Control Regions & $\ttbar$+$\PZ$ & \PW\PZ & \PZ{}\PZ  & Fake $\Pgt$\\
\midrule
 
Observed events   & 67 & 96 & 774 & 7743 \\
\hline
 
Total background  & $67 \pm 8\phantom{0}$ & $97 \pm 9\phantom{0}$ & $774 \pm 28\phantom{0}$ & $7760 \pm 90\phantom{00}$ \\
\hline
$Z+$jets   &  $<0.03$ & $0.1 \pm 0.1$ & $<0.03$ & $2.55 \pm 0.32$ \\
Other Top   & $6.3 \pm 0.9$ & $2.0 \pm 0.6$ & $0.73 \pm 0.14$ & $66 \pm 7\phantom{0}$ \\
$t\bar{t}$+$Z$   & $51 \pm 8\phantom{0}$ & $3.1 \pm 0.6$ & $6.5 \pm 1.2$ & $52 \pm 9\phantom{0}$ \\
$ZZ$   & $4.8 \pm 0.5$ & $2.5 \pm 0.2$ & $753 \pm 28\phantom{0}$ & $95 \pm 5\phantom{0}$ \\
$WZ$  & $10.0 \pm 0.1\phantom{0}$ & $82 \pm 9\phantom{0}$ & $0.17 \pm 0.03$ & $370 \pm 50\phantom{0}$ \\
Triboson   &   $<0.03$ & $0.9 \pm 0.1$ & $0.44 \pm 0.04$ & $1.71 \pm 0.15$  \\
Fakes   & $4.8 \pm 0.8$ & $7.0 \pm 0.7$ & $12.4 \pm 1.2\phantom{0}$  & $7170 \pm 100\phantom{0}$ \\
\hline
$\tau'$(\SI{130}{\GeV})*    & $11 \pm 4\phantom{0}$ & $12 \pm 4\phantom{0}$ & $285 \pm 21\phantom{0}$ & $1380 \pm 100\phantom{0}$ \\
$\tau'$(\SI{500}{\GeV})*    & $0.7 \pm 0.1$ & $7.4 \pm 0.5$ & $0.64 \pm 0.13$ & $3.71 \pm 0.31$  \\
$\tau'$(\SI{800}{\GeV})*    &  $<0.03$ & $3.1 \pm 0.2$ &  $<0.03$ & $0.06 \pm 0.02$   \\
$\tau'$(\SI{1000}{\GeV})*    &  $<0.03$ & $1.2 \pm 0.1$ &  $<0.03$ &  $<0.03$   \\
 
\bottomrule
\multicolumn{5}{l}{*Pre-fit}
\end{tabular}
\label{tab:yieldsCR}
\end{center}
\end{table}


Table~\ref{tab:rank} shows the
impact of each source of systematic uncertainty on the signal strength $\mu$ in
the signal-plus-background fit. Signal strength is defined as the ratio of the signal cross section estimated
using the data to the predicted signal cross section. The nuisance parameters are grouped according to their origin.
To evaluate the impact of each source of systematic uncertainty, the source is removed from the full fit and the
signal strength and its uncertainty are recalculated.
The square of the impact is defined as the decrease in the squared signal-strength uncertainty.
The nuisance parameters associated with the background normalization have the highest
impact on $\mu$, while the systematic uncertainties associated with the fake-lepton estimation have the second-highest impact.

\begin{table}[htbp]
\caption{Impact of each source of systematic uncertainty on the signal strength $\mu$ in the signal-plus-background fit. $M_{\tau'}=900$~GeV is considered.
Both the positive and negative impacts on $\mu$ are considered. Individual sources of uncertainty are grouped for brevity but are treated as independent and uncorrelated in the fitting procedure. Impact of the total statistical uncertainty is also presented.}
\begin{center}
\begin{tabular}{l|ll}
\toprule
Uncertainty&$+\Delta \mu$&$-\Delta \mu$\\
\midrule
Normalization factors & 0.046&0.059\\
Fakes&0.033&0.050\\
Other theory uncertainties& 0.017&0.012\\
$\tau$-lepton ID and reconstruction&0.016&0.045\\
Light-lepton ID, reco., energy scale and momentum resolution&0.016&0.033\\
Jet energy scale and resolution, JVT, \pileup reweighting&0.016&0.012\\
$\tau$-lepton energy scale&0.011&0.010\\
\MET &0.008&0.003\\
Luminosity & 0.015&0.006\\
Flavour tagging&0.003&0.003\\
MC and fake-background statistical uncertainty & 0.066&0.120\\
\hline
Total systematic uncertainty & 0.098&0.160\\
\hline
Total statistical uncertainty & 0.450&0.350\\
\bottomrule
\end{tabular}
\label{tab:rank}
\end{center}
\end{table}
 
Using the VLL doublet model in Refs.~\cite{Kumar:2015tna, Bhattiprolu:2019vdu}, the predicted
significance of the signal is expected to be greater than 5~standard deviations for \Mtaup
in the range from \SIrange[range-units=single]{130}{600}{\GeV} and greater than 3~standard deviations
for values of
\Mtaup up to \SI{800}{\GeV}.
Above this mass the expected significance decreases to $\approx 1$~standard
deviation at
\SI{1}{\TeV}. The observed significance is found to be
$\lessapprox 1$~standard deviation over the
entire \Mtaup range probed, as shown in Table~\ref{tab:results}.
 
\begin{table}[htbp]
\caption{Results from the fitting procedure for the VLL mass points studied. The expected and observed signal significances, and the 95\% CL exclusion limits are shown.}
\begin{center}
\begin{tabular}{l |
S[table-format=2.2,table-figures-decimal=1,table-auto-round] |
S[table-format=1.2,table-figures-decimal=1,table-auto-round] |
S[table-format=4.2,table-figures-decimal=1,table-auto-round] |
S[table-format=1.2,table-figures-decimal=1,table-auto-round]}
\toprule
$\tau'$ Mass [GeV]&\multicolumn{2}{c}{Significance}&\multicolumn{2}{| c}{Exclusion Limit [fb]}\\
\midrule
&\parbox{.1\textwidth}{\centering Expected}&\parbox{.15\textwidth}{\centering Observed}&\parbox{.15\textwidth}{\centering Expected}&\parbox{.15\textwidth}{\centering Observed}\\
\midrule
 
\phantom{1}130&5.89&-0.44& \parbox{.15\textwidth}{\centering $\phantom{.^0}1110^{+520}_{-310}\phantom{00^{.}}$} &953\\[0.5ex]
\phantom{1}200&12&-0.41& \parbox{.15\textwidth}{\centering $\phantom{.}100^{+44}_{-28}\phantom{0^{.}}$} &90\\[0.5ex]
\phantom{1}300&15&0.09& \parbox{.15\textwidth}{\centering $17.0^{+7.5}_{-4.8}\phantom{0}$} &18\\[0.5ex]
\phantom{1}400&12&0.31& \parbox{.15\textwidth}{\centering $7.4^{+3.3}_{-2.1}$} &8.66\\[0.5ex]
\phantom{1}500&10&0.69& \parbox{.15\textwidth}{\centering $3.6^{+1.6}_{-1.0}$} &4.69\\[0.5ex]
\phantom{1}600&7.57&0.91& \parbox{.15\textwidth}{\centering $2.3^{+1.0}_{-0.6}$} &3.11\\[0.5ex]
\phantom{1}700&5.06&0.96& \parbox{.15\textwidth}{\centering $1.8^{+0.8}_{-0.5}$} &2.44\\[0.5ex]
\phantom{1}800&3.35&1.10& \parbox{.15\textwidth}{\centering $1.5^{+0.7}_{-0.4}$} &2.08\\[0.5ex]
\phantom{1}900&2.14&1.11& \parbox{.15\textwidth}{\centering $1.3^{+0.6}_{-0.4}$} &1.82\\[0.5ex]
1000&1.34&1.19& \parbox{.15\textwidth}{\centering $1.2^{+0.6}_{-0.3}$} &1.74\\[0.5ex]
1100&0.79&1.12& \parbox{.15\textwidth}{\centering $1.2^{+0.6}_{-0.3}$} &1.72\\[0.5ex]
1200&0.47&1.12& \parbox{.15\textwidth}{\centering $1.2^{+0.6}_{-0.3}$} &1.72\\[0.5ex]
1300&0.28&1.14& \parbox{.15\textwidth}{\centering $1.2^{+0.6}_{-0.3}$} &1.72\\[0.5ex]
\bottomrule
\end{tabular}
\label{tab:results}
\end{center}
\end{table}

\begin{figure}[hbtp]
\centering
\includegraphics[width=0.70\textwidth]{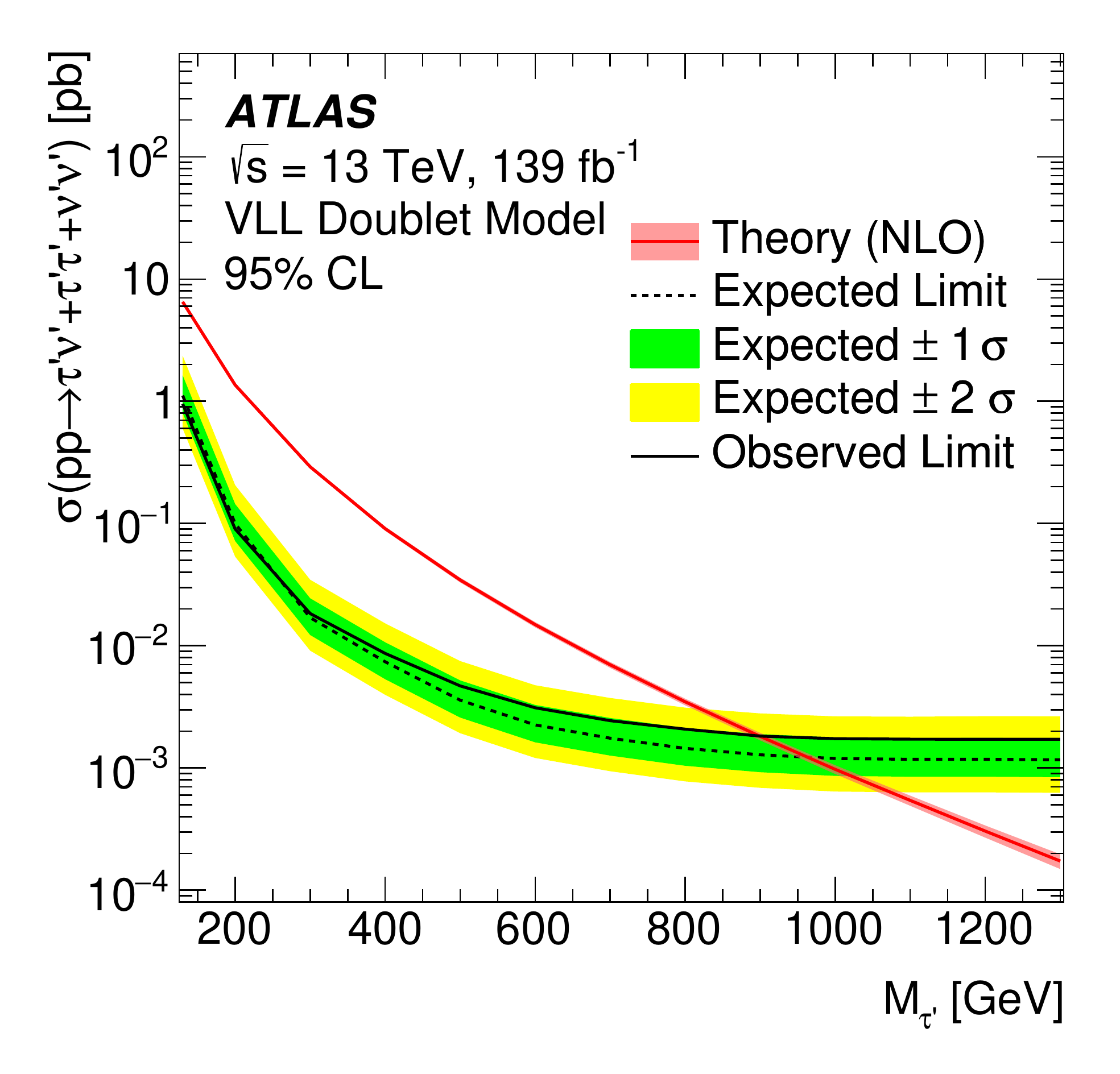}
\caption{The 95\% CL exclusion limit on the VLL production cross section as a function of VLL mass. The black dashed line represents the expected limit while the shaded regions are its one and two~standard-deviation uncertainty bands. The solid black line is the observed limit as a function of VLL mass. The red curve is the NLO theory prediction along with its uncertainty. \label{fig:ExpectedLimit}
}
\end{figure}

No significant deviation from the SM prediction is observed. Therefore,
the 95\% CL exclusion limit on the VLL production cross section as a function of \Mtaup is
calculated and shown in Figure~\ref{fig:ExpectedLimit}. In order to estimate the 95\% CL upper limit on the VLL cross section,
a simultaneous binned likelihood fit of the seven SRs and four CRs is performed,
using the \cls method~\cite{Read:2002hq} with the asymptotic approximation.
The expected limit is shown with the black dashed line, and the shaded regions correspond to its one and two~standard-deviation uncertainty bands.
The observed 95\% CL exclusion limit is shown with the solid black line.
The expected lower limit on \Mtaup is found to be \SI{970}{\GeV}, and the observed limit to be \SI{900}{GeV}, by comparing the NLO theory prediction with the expected and observed 95\% CL cross-section limits.
 
\FloatBarrier


\section{Conclusions}
\label{sec:conclusion}

A search for vector-like leptons in a doublet model is performed
using \SI{139}{\ifb} of \pp collision data recorded at $\sqs=\SI{13}{\TeV}$
by the ATLAS detector at the LHC. The search is performed using events
with final states containing multiple light leptons and \tauhad. Observing
no excess of events above the SM expectation, a 95\% CL upper limit
on the cross section is calculated using the \cls method.
Using a doublet model where the vector-like
leptons couple to the third-generation SM leptons, the observed
mass range from \SIrange[range-units=single]{130}{900}{\GeV} is excluded at the 95\% CL, while the
highest excluded mass is expected to be \SI{970}{\GeV}.


\section*{Acknowledgements}


We thank CERN for the very successful operation of the LHC, as well as the
support staff from our institutions without whom ATLAS could not be
operated efficiently.
 
We acknowledge the support of
ANPCyT, Argentina;
YerPhI, Armenia;
ARC, Australia;
BMWFW and FWF, Austria;
ANAS, Azerbaijan;
CNPq and FAPESP, Brazil;
NSERC, NRC and CFI, Canada;
CERN;
ANID, Chile;
CAS, MOST and NSFC, China;
Minciencias, Colombia;
MEYS CR, Czech Republic;
DNRF and DNSRC, Denmark;
IN2P3-CNRS and CEA-DRF/IRFU, France;
SRNSFG, Georgia;
BMBF, HGF and MPG, Germany;
GSRI, Greece;
RGC and Hong Kong SAR, China;
ISF and Benoziyo Center, Israel;
INFN, Italy;
MEXT and JSPS, Japan;
CNRST, Morocco;
NWO, Netherlands;
RCN, Norway;
MEiN, Poland;
FCT, Portugal;
MNE/IFA, Romania;
MESTD, Serbia;
MSSR, Slovakia;
ARRS and MIZ\v{S}, Slovenia;
DSI/NRF, South Africa;
MICINN, Spain;
SRC and Wallenberg Foundation, Sweden;
SERI, SNSF and Cantons of Bern and Geneva, Switzerland;
MOST, Taiwan;
TENMAK, T\"urkiye;
STFC, United Kingdom;
DOE and NSF, United States of America.
In addition, individual groups and members have received support from
BCKDF, CANARIE, Compute Canada and CRC, Canada;
PRIMUS 21/SCI/017 and UNCE SCI/013, Czech Republic;
COST, ERC, ERDF, Horizon 2020 and Marie Sk{\l}odowska-Curie Actions, European Union;
Investissements d'Avenir Labex, Investissements d'Avenir Idex and ANR, France;
DFG and AvH Foundation, Germany;
Herakleitos, Thales and Aristeia programmes co-financed by EU-ESF and the Greek NSRF, Greece;
BSF-NSF and MINERVA, Israel;
Norwegian Financial Mechanism 2014-2021, Norway;
NCN and NAWA, Poland;
La Caixa Banking Foundation, CERCA Programme Generalitat de Catalunya and PROMETEO and GenT Programmes Generalitat Valenciana, Spain;
G\"{o}ran Gustafssons Stiftelse, Sweden;
The Royal Society and Leverhulme Trust, United Kingdom.
 
The crucial computing support from all WLCG partners is acknowledged gratefully, in particular from CERN, the ATLAS Tier-1 facilities at TRIUMF (Canada), NDGF (Denmark, Norway, Sweden), CC-IN2P3 (France), KIT/GridKA (Germany), INFN-CNAF (Italy), NL-T1 (Netherlands), PIC (Spain), ASGC (Taiwan), RAL (UK) and BNL (USA), the Tier-2 facilities worldwide and large non-WLCG resource providers. Major contributors of computing resources are listed in Ref.~\cite{ATL-SOFT-PUB-2023-001}.


\printbibliography
 
\clearpage
 
\begin{flushleft}
\hypersetup{urlcolor=black}
{\Large The ATLAS Collaboration}

\bigskip

\AtlasOrcid[0000-0002-6665-4934]{G.~Aad}$^\textrm{\scriptsize 101}$,
\AtlasOrcid[0000-0002-5888-2734]{B.~Abbott}$^\textrm{\scriptsize 119}$,
\AtlasOrcid[0000-0002-7248-3203]{D.C.~Abbott}$^\textrm{\scriptsize 102}$,
\AtlasOrcid[0000-0002-1002-1652]{K.~Abeling}$^\textrm{\scriptsize 55}$,
\AtlasOrcid[0000-0002-8496-9294]{S.H.~Abidi}$^\textrm{\scriptsize 29}$,
\AtlasOrcid[0000-0002-9987-2292]{A.~Aboulhorma}$^\textrm{\scriptsize 35e}$,
\AtlasOrcid[0000-0001-5329-6640]{H.~Abramowicz}$^\textrm{\scriptsize 150}$,
\AtlasOrcid[0000-0002-1599-2896]{H.~Abreu}$^\textrm{\scriptsize 149}$,
\AtlasOrcid[0000-0003-0403-3697]{Y.~Abulaiti}$^\textrm{\scriptsize 116}$,
\AtlasOrcid[0000-0003-0762-7204]{A.C.~Abusleme~Hoffman}$^\textrm{\scriptsize 136a}$
% ...
\bigskip
\\

% ...
$^{29}$Physics Department, Brookhaven National Laboratory, Upton NY; United States of America.\\
% ...
$^{35}$$^{(a)}$Facult\'e des Sciences Ain Chock, R\'eseau Universitaire de Physique des Hautes Energies - Universit\'e Hassan II, Casablanca;$^{(b)}$Facult\'{e} des Sciences, Universit\'{e} Ibn-Tofail, K\'{e}nitra;$^{(c)}$Facult\'e des Sciences Semlalia, Universit\'e Cadi Ayyad, LPHEA-Marrakech;$^{(d)}$LPMR, Facult\'e des Sciences, Universit\'e Mohamed Premier, Oujda;$^{(e)}$Facult\'e des sciences, Universit\'e Mohammed V, Rabat;$^{(f)}$Institute of Applied Physics, Mohammed VI Polytechnic University, Ben Guerir; Morocco.\\
% ...
$^{55}$II. Physikalisches Institut, Georg-August-Universit\"{a}t G\"ottingen, G\"ottingen; Germany.\\
% ...
$^{101}$CPPM, Aix-Marseille Universit\'e, CNRS/IN2P3, Marseille; France.\\
$^{102}$Department of Physics, University of Massachusetts, Amherst MA; United States of America.\\
% ...
$^{116}$Department of Physics, New York University, New York NY; United States of America.\\
% ...
$^{119}$Homer L. Dodge Department of Physics and Astronomy, University of Oklahoma, Norman OK; United States of America.\\
% ...
$^{136}$$^{(a)}$Departamento de F\'isica, Pontificia Universidad Cat\'olica de Chile, Santiago;$^{(b)}$Millennium Institute for Subatomic physics at high energy frontier (SAPHIR), Santiago;$^{(c)}$Instituto de Investigaci\'on Multidisciplinario en Ciencia y Tecnolog\'ia, y Departamento de F\'isica, Universidad de La Serena;$^{(d)}$Universidad Andres Bello, Department of Physics, Santiago;$^{(e)}$Instituto de Alta Investigaci\'on, Universidad de Tarapac\'a, Arica;$^{(f)}$Departamento de F\'isica, Universidad T\'ecnica Federico Santa Mar\'ia, Valpara\'iso; Chile.\\
% ...
$^{149}$Department of Physics, Technion, Israel Institute of Technology, Haifa; Israel.\\
$^{150}$Raymond and Beverly Sackler School of Physics and Astronomy, Tel Aviv University, Tel Aviv; Israel.\\
% ...
$^{a}$ Also Affiliated with an institute covered by a cooperation agreement with CERN.\\
$^{b}$ Also at An-Najah National University, Nablus; Palestine.\\
$^{c}$ Also at Borough of Manhattan Community College, City University of New York, New York NY; United States of America.\\
$^{d}$ Also at Bruno Kessler Foundation, Trento; Italy.\\
$^{e}$ Also at Center for High Energy Physics, Peking University; China.\\
$^{f}$ Also at Centro Studi e Ricerche Enrico Fermi; Italy.\\
% ...

\end{flushleft}


\end{document}